\documentclass{imsart}

\usepackage{natbib}
\usepackage{comment}
\usepackage{hyperref}

\usepackage{amsmath,bm,amsfonts,color,amsthm}
\usepackage{graphicx}
\usepackage[margin=1.4in,footskip=0.25in]{geometry}
\usepackage{fancyhdr,graphicx,amsmath,amssymb}

\usepackage{caption}
\usepackage{subcaption}

\usepackage{float}
\restylefloat{table}

\usepackage{tikz}
\usetikzlibrary{fit,positioning,calc}
\usepackage{algorithm}
\usepackage{algorithmic}
\usepackage{float}

\makeatletter
\newenvironment{breakablealgorithm}
  {
   \begin{center}
     \refstepcounter{algorithm}
     \hrule height.8pt depth0pt \kern2pt
     \renewcommand{\caption}[2][\relax]{
       {\raggedright\textbf{\ALG@name~\thealgorithm} ##2\par}%
       \ifx\relax##1\relax 
         \addcontentsline{loa}{algorithm}{\protect\numberline{\thealgorithm}##2}%
       \else 
         \addcontentsline{loa}{algorithm}{\protect\numberline{\thealgorithm}##1}%
       \fi
       \kern2pt\hrule\kern2pt
     }
  }{
     \kern2pt\hrule\relax
   \end{center}
  }
\makeatother
\def\T{{ \mathrm{\scriptscriptstyle T} }}

\renewcommand{\epsilon}{\varepsilon}

\let\originalleft\left
\let\originalright\right
\renewcommand{\left}{\mathopen{}\mathclose\bgroup\originalleft}
\renewcommand{\right}{\aftergroup\egroup\originalright}

\def\bas#1\eas{\begin{align*}#1\end{align*}}
\def\basn#1\easn{\begin{align}#1\end{align}}

\definecolor{dg}{RGB}{0,170,0}

\graphicspath{{images/}}

\begin{document}
\begin{frontmatter}
\title{Bayesian modeling of networks \\ in complex business intelligence problems}
\runtitle{Modeling of complex network data in marketing}
\begin{aug}
\author{\fnms{Daniele} \snm{Durante}\thanksref{m1}\ead[label=e1]{durante@stat.unipd.it}},
  \author{\fnms{Sally} \snm{Paganin}\thanksref{m1}\ead[label=e2]{sallypaganin@gmail.com}},
   \author{\fnms{Bruno} \snm{Scarpa}\thanksref{m1}\ead[label=e3]{scarpa@stat.unipd.it}} 
   \and
\author{\fnms{David B.} \snm{Dunson}\thanksref{m2}\ead[label=e4]{dunson@duke.edu}}
  \runauthor{Durante, Paganin, Scarpa and Dunson}
      \affiliation{University of Padua}
   \address{\thanksmark{m1}University of Padua, Department of Statistical Sciences, \\Via Cesare Battisti 241, 35121 Padua, Italy\\ \printead{e2}, \printead{e1}, \printead{e3}}
   \address{\thanksmark{m2}Duke University, Department of Statistical Science,\\
Box 90251, Durham, NC 27708 34127, USA.\\
\printead{e4}}
   \end{aug}

\begin{abstract}
Complex network data problems are increasingly common in many fields of application. Our motivation is drawn from strategic marketing studies monitoring customer choices of specific products, along with co-subscription networks encoding multiple purchasing behavior. Data are available for several agencies within the same insurance company, and our goal is to efficiently exploit co-subscription networks to inform targeted advertising of cross-sell strategies to currently mono-product customers.  We address this goal by developing a Bayesian hierarchical model, which clusters agencies according to common mono-product customer choices and co-subscription networks. Within each cluster, we efficiently model customer behavior via a cluster-dependent mixture of latent eigenmodels. This formulation provides key information on mono-product customer choices and multiple purchasing behavior within each cluster, informing targeted cross-sell strategies.  We develop simple algorithms for tractable inference, and assess performance in simulations and an application to business intelligence.
\end{abstract}

\end{frontmatter}
\section{Introduction}\label{sec_1}
Increasing business competition and market saturation have led companies to progressively shift the focus of their marketing strategies from the acquisition of new customers to an increased penetration of their customer base. Targeting existing customers via cross-sell campaigns instead of attracting new ones provides a more effective strategy for the growth of the company and enhances customer retention by increasing the switching costs  \citep{kama_1991}. Therefore, mono-product customers purchasing a single product from a company represent a key segment of the customer base, and companies are naturally interested in expanding these customers purchases to additional products.

Business statistics currently offers a wide set of procedures for cross-sell relying on shared acquisition patterns of products, based on customer ownership data. A first effort in addressing this aim can be found in the latent trait model developed by \citet{kama_1991} to estimate the propensity of a customer towards a particular product, based on its ownership of other products.  This procedure has been later improved by \citet{kama_2003} combining information from customer databases with survey data. \citet{ver_2001} focus instead on predicting the potential value of a current customer via a multivariate probit model, and propose a two-by-two segmentation to improve targeting. Finally, \citet{tur_2012} develops a multivariate credibility method to identify a profitable set of customers for cross-selling. This is accomplished by estimating a latent risk profile for each customer, exploiting information on claims. Refer also to \citet{tur_2012a} and \citet{kai_2013} for recently developed cross-sell strategies and for an overview on available methodologies.

Previous procedures exploit different sources, including customer demographics and survey data, to estimate co-subscription probabilities among pairs of products for each customer in a single agency. Differently from this setting, we do not observe customer demographic data for a single agency, but monitor mono-product customer choices along with co-subscription networks among $V=15$ products  for $n=130$ agencies operating in the Italian insurance market. Customer relationship management is becoming increasingly important to effectively operate in the insurance market. This sector is mostly stable in developed countries, and rising customer expectations, along with tight competition among top corporations and low growth potentials, force companies to efficiently exploit their databases to create, manage and maintain their portfolio of profitable customers \citep{mati_2014}.

\begin{figure}[t]
\centering
\includegraphics[height=9.6cm, width=15cm]{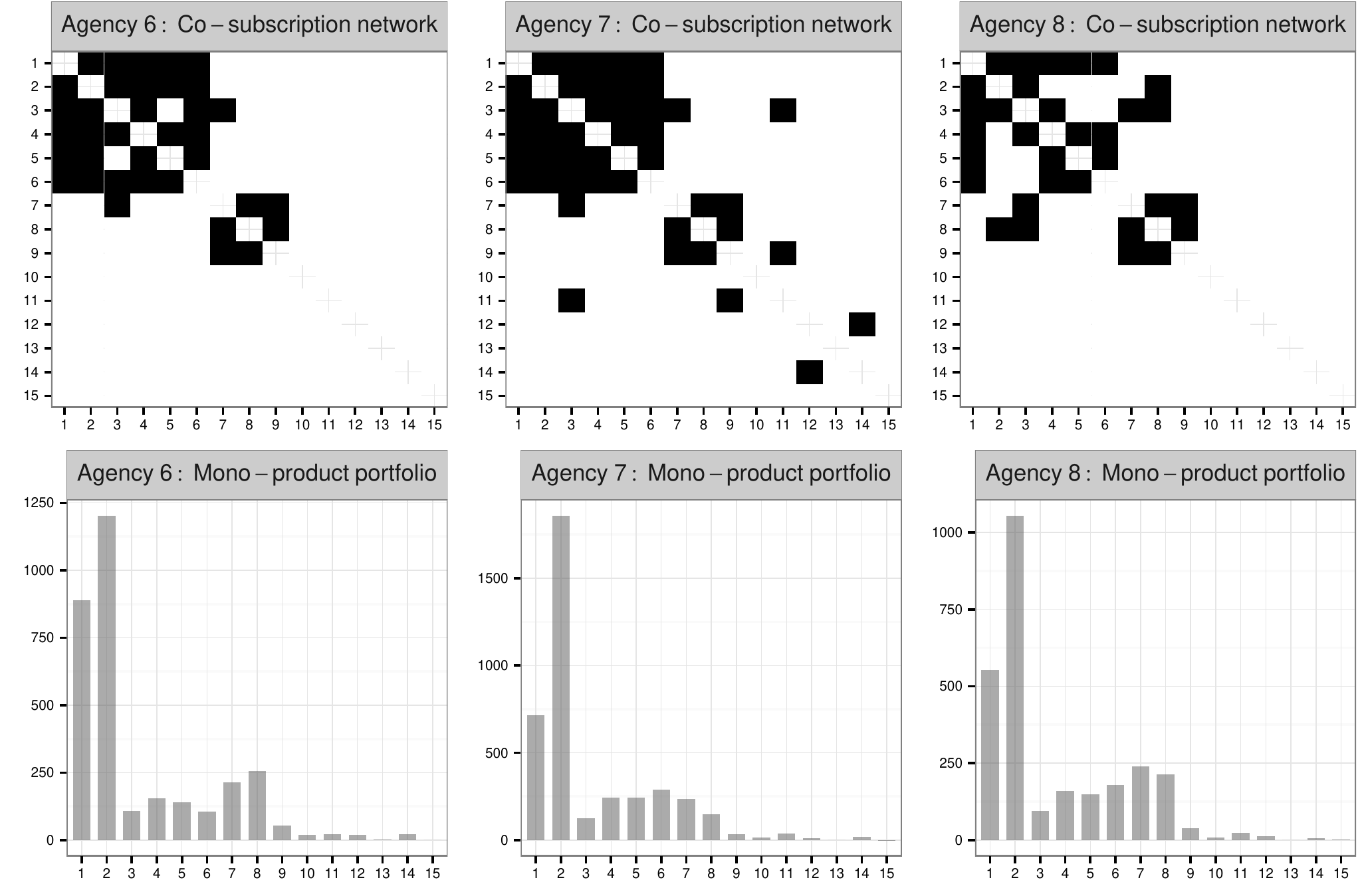}
\caption{\footnotesize{For three selected agencies. Upper panel: observed co-subscription networks $A_i$. Black refers to an edge, white to a non-edge. Lower panel: total number of mono-product customers for each product $v=1, \ldots, 15$ based on choice data $y_{is}$, $s=1, \ldots, n_i$.}}
\label{F1}
\end{figure}

We observe choice data $y_{is} \in \{1, \ldots, V \}$ denoting the product subscribed to by mono-product customer $s=1,\ldots,n_i$, within agency $i=1,\ldots,n$. The co-subscription network for each agency $i=1, \ldots, n$ is  available via a $V \times V$ symmetric adjacency matrix $A_i$, with $A_{i[vu]}=A_{i[uv]}=1$ if -- in agency $i$ -- the number of customers subscribed to both products $v$ and $u$, exceeds $10\%$ of the total amount of multi-product customers subscribed to at least one of the two, for every $v=2, \ldots, V$ and $u=1, \ldots, v-1$.  A presence of an edge between two products suggests a preference of customers in agency $i$ for that specific pair, controlling for the total number of multi-product customers subscribed to at least one of the two products. Refer to Figure  \ref{F1} for an example of available data.

Although the $10\%$ threshold can be adapted to the company requirements, we found the resulting adjacency matrices $A_1, \ldots, A_{130}$ robust to moderate changes in this value. In fact, when the threshold is set at $7.5\%$ instead of $10\%$, the resulting adjacency matrices differ from the previous ones for only $6\%$ of the pairs. This gap remains on similarly low values $4\%$ and $8\%$ when the threshold is set at $12.5\%$ and $15\%$, respectively. 

Each agency can define appropriate cross-sell strategies by exploiting its co-subscription network $A_i$ to estimate the propensity of a customer who subscribed to product $v=1, \ldots, V$ to additionally buy $u \neq v$. This leads to $V$ different cross-sell strategies $u_{i1}, \ldots, u_{iV}$, with $u_{iv}$ defining which additional product  $u \neq v$ is the best offer to currently mono-product customers subscribed to $v$ in agency $i$. Hence, $u_{iv}=\mbox{argmax}_u\{ \mbox{pr}(\mathcal{A}_{i[vu]}=1): u \neq v\}$, with $\mathcal{A}_{i[vu]}$ the random variable characterizing the presence or absence of an edge between products $v$ and $u$ in agency $i$. Efficiently targeting advertising by offering customers the product mostly complementary to their current choice can substantially improve performance relative to untargeted advertising, while increasing the satisfaction and reducing churn effects due to frequent and pointless cross-selling attempts  \citep{azza_2012,kama_2003}.  Satisfied customers are a key to enhancing positive word-of-mouth communication and are less sensitive to competing brands and price  \citep{mati_2014}.

The effectiveness at increasing the number of multi-product customers in agency $i$ depends not only on the propensity of customers with $v$ to also subscribe to $u$, measured by $\mbox{pr}(\mathcal{A}_{i[vu]}=1)$, but also on the proportion of mono-product customers with $v$, defined by $p_{iv}=\mbox{pr}(\mathcal{Y}_{i}=v)$, with $\mathcal{Y}_{i}$ the random variable denoting the choices of mono-product customers in agency $i$. If $p_{iv}$ is low, then strategy $u_{iv}$ targets a small portion of the customer base of agency $i$, and hence has a low ceiling on effectiveness. To take into account the role of $p_{iv}$, we associate each strategy $u_{iv}$ with a performance indicator $e_{iv}=p_{iv} \mbox{max}\{\mbox{pr}(\mathcal{A}_{i[vu]}=1): u \neq v\}$, for each $v=1, \ldots, V$ and $i=1, \ldots, n$.  Strategies with a high $e_{iv}$ will target a sizable proportion of the available customers for that agency with advertising for a new product likely to be appealing to them.

In defining and evaluating cross-sell strategies, there are two important issues to take into consideration.  Firstly, we are faced with statistical error in estimating the components underlying strategies $u_{iv}$ and indicators $e_{iv}$, for each $v=1, \ldots, V$ and $i=1, \ldots, n$. This is a particular problem in estimating $\mbox{pr}(\mathcal{A}_{i[vu]}=1)$ due to data sparsity.  The second issue is that it is important to take into account the fact that administrative overhead can be reduced by considering the same strategy for different agencies within the same company.  For groups of agencies having sufficiently similar customer bases, an identical strategy can be adopted to reduce administrative costs without decreasing effectiveness.  Motivated by this notion, we propose to address both the statistical error and administrative overhead issue through clustering of agencies according to the parameters characterizing their customer bases, and then administrating the same strategy to all agencies within a cluster.  

As suggested by Figure  \ref{F1},  it is reasonable to expect agencies operating in similar markets to exhibit clusters, corresponding to common patterns in the composition of their  mono-product customer choices and co-subscription networks. Efficient detection of such clusters allows adaptive reduction of the total number of strategies to be devised from $u_{i1}, \ldots, u_{iV}$, $i=1, \ldots, n$  to $u_{k1}, \ldots, u_{kV}$, $k=1, \ldots, K<n$, with each cluster-specific set of strategies  $u_{k1}, \ldots, u_{kV}$ maintaining its effectiveness in targeting similar agencies. This higher level targeting and profiling represents a key to balance the need of the company to reduce costs and the importance of providing agencies with effective strategies that account for their specific structure. Providing agencies with sets of strategies suitably related with their structure is further important to increase their trust in the company and improve synergy.  

We address this goal by developing a Bayesian hierarchical model, which adaptively associates shared sets of strategies $u_{k1}, \ldots, u_{kV}$ and performance indicators $e_{k1}, \ldots, e_{kV}$ to clusters of agencies characterized by mono- and multi-product customers with similar purchasing behavior. Each cluster-specific set of cross-sell strategies $u_{k1}, \ldots, u_{kV}$ is devised by learning cluster-specific co-subscription patterns among pairs of products, exploiting data on multi-product customers. Joining this information with the estimated cluster-specific distribution of 
mono-product customers across products, performance indicators  $e_{k1}, \ldots, e_{kV}$  are constructed.  To our knowledge, this is the first approach in the literature that considers a two-level cross-sell segmentation of the customer base, which clusters agencies with similar customer preferences, and profiles mono- and multi-product purchasing behavior within each cluster to define cross-sell strategies and related performance indicators.

\subsection{Joint modeling of mixed domain data}
As a step towards our goal of designing efficient cross-sell strategies, we first develop a joint model for the data $\{{\bf y}_i=(y_{i1}, \ldots, y_{in_i}),A_i\}$, $i=1, \ldots, n$, which characterizes the distribution across agencies of the mono-product customer subscriptions along with the co-subscription network for multi-product customers.  The model is chosen to be flexible while automatically clustering different agencies that have similar mono-product customer choices and co-subscription network profiles.  This clustering is useful for borrowing information across agencies in efficiently and effectively learning the joint distribution of the mono- and multi-product purchasing behavior of the customer bases.  In addition, clustering provides a simplification, which is useful in design of interpretable strategies.

There is an increasing statistical literature on joint modeling and co-clustering of mixed domain data. Available procedures focus on learning the association between a univariate response variable and an object predictor, typically characterized by a function. \citet{big_2009}  favor clustering among predictor trajectories, with each cluster associated to a specific offset in a generalized linear model for the response variable. Although providing an appealing procedure for the sake of interpretability and inference, their model lacks flexibility in constraining predictor and response clusters to be the same. This may require the introduction of many clusters to appropriately characterize the joint distribution of the mixed domain data, reducing the performance in estimating cluster-specific components and providing a biased overview of the underlying clustering structure. 

\citet{duns_2008} address the previous issue by modeling the conditional distribution of the response via a cluster-dependent mixture representation, rather than considering only a cluster-specific offset in the conditional expectation. In improving flexibility via dependent mixture modeling, their procedure can better identify the underlying clustering structure; see also \citet{bane_2013} for a recent overview of this topic. Although we are similar to the previous methods in looking for flexible and accurate joint modeling and co-clustering procedures for mixed domain data, our motivating data set is substantially different in considering categorical mono-product customer choices and network-valued co-subscription data. Flexible modeling of the conditional distribution of a network-valued random variable is still an ongoing issue, which requires careful representations in order to borrow information across edges, reduce the dimensionality and maintain flexibility. 

We propose a cluster-dependent mixture of latent eigenmodels, which allows the distribution of the co-subscription networks to flexibly change across clusters of agencies via cluster-specific mixing probabilities, while borrowing information among these agencies in learning the shared mixture components. Considering cluster dependence only in the mixing probabilities allows further dimensionality reduction, while providing simple and efficient computational methods. Differently from \citet{duns_2008}, we additionally avoid fixing the total number of clusters, but learn this key quantity from our data via  a Chinese Restaurant Process prior for the cluster assignments.

The paper is organized as follows. In Section \ref{sec_2}, we carefully describe our hierarchical formulation for joint modeling and co-clustering of co-subscription networks and mono-product choice data. Prior specification and detailed steps for posterior computation are outlined in Section \ref{sec_3}. Section \ref{sec_4} considers simulation studies to evaluate the performance of our model in a scenario mimicking our motivating application. Results of the application to our business intelligence problem in an insurance company are presented and discussed in Section \ref{sec_5}. Finally, Section \ref{sec_6} contains concluding remarks and suggests other possible fields of application along with further research directions.

\section{Joint modeling of mono-product choices and co-subscription networks}\label{sec_2}
Let ${\bf C}=(C_1,\ldots,C_n)$ denote a vector of cluster assignments, with $C_i \in \{1, \ldots,  K\}$ indicating the cluster membership of agency $i$. Agencies within the same cluster are characterized by a similar composition of their mono-product customer choices as well as a comparable co-subscription network. To complete a specification of the joint model, we need to define a cluster-specific probabilistic representation of the mono-product customer choices, as well as a cluster-specific probabilistic generative mechanism underlying the co-subscription networks. The latter is a key to define cross-sell strategies $u_{k1}, \ldots, u_{kV}$ in each cluster $k=1, \ldots, K$, while the former provides the additional information to define the performance indicators $e_{k1}, \ldots, e_{kV}$.

As a mono-product customer can be associated with only one subscription $v=1, \ldots, V$, it is straightforward to define a probabilistic representation for the mono-product customer choices within each cluster. Let $\mathcal{Y}_k$ denote the categorical random variable characterizing the choices of the mono-product customers in agencies within cluster $k$, for each $k=1, \ldots, K$. To define the probability mass function associated to each $\mathcal{Y}_k$, we  introduce a cluster-specific vector ${\bf p}_k=(p_{k1}, \ldots, p_{kV})$, with $p_{kv}$ indicating the probability that a mono-product customer in an agency within cluster $k$ subscribes to product $v$, for each $v=1, \ldots, V$.  Assuming independence of the mono-product customer choices, the joint probability for data ${\bf y}_i$ in agency $i$ given its membership to cluster $k$ is
\begin{eqnarray}
\mbox{pr}(\mathcal{Y}_k=y_{i1})\mbox{pr}(\mathcal{Y}_k=y_{i2}) \cdots \mbox{pr}(\mathcal{Y}_k=y_{in_i})= \prod_{v=1}^V p_{kv}^{n_{iv}}, 
\label{eq_1}
\end{eqnarray}
with $n_{iv}$ the number of mono-product customers in agency $i$ who subscribed to product $v$.  

Within each cluster $k$, co-subscription networks  are realizations from a network-valued random variable. As our network data are undirected and self-relations are not of interest, each symmetric adjacency matrix $A_i$, $i=1, \ldots, n$, is uniquely characterized by its lower triangular elements comprising ${\mathcal{L}}({A}_i)=(A_{i[21]}, A_{i[31]}, \ldots, A_{i[V1]}, A_{i[32]}, \ldots,  A_{i[V2]}, \ldots, A_{i[V(V-1)]})$. Hence, in defining a probabilistic generative mechanism for the co-subscription networks within each cluster $k$, we can focus on the multivariate random variable ${\mathcal{L}}({\mathcal{A}}_k)$ with binary elements $\mathcal{L}(\mathcal{A}_{k})_l \in \{0,1\}$,  measuring the presence or absence of an edge among each pair of products $l=1, \ldots, V(V-1)/2$ for agencies in cluster $k$. Note that in our notation ${\mathcal{L}}(\cdot)$ is an operator vectorizing the lower triangular elements of a given symmetric matrix.

As there are $2^{V(V-1)/2}$ possible configurations of co-subscription networks among $V$ products, we cannot estimate the probability mass function associated to each ${\mathcal{L}}({\mathcal{A}}_k)$, $k=1, \ldots, K$ nonparametrically without dimensionality reduction.  To reduce dimension while maintaining flexibility, we model each ${\mathcal{L}}({\mathcal{A}}_k)$ via a  cluster-dependent mixture of latent eigenmodels. This leads to the following probability for the co-subscription network $A_i$ in agency $i$ given its membership to cluster $k$:
\begin{eqnarray}
\mbox{pr}\{{\mathcal{L}}({\mathcal{A}}_k)={\mathcal{L}}({{A}}_i)\}=  \sum_{h=1}^{H} \nu_{hk} \prod_{l=1}^{V(V-1)/2} \left\{\pi_{l}^{(h)}\right\}^{\mathcal{L}({{A}}_i)_l} \left\{1-  \pi^{(h)}_{l}\right\}^{1-\mathcal{L}({{A}}_i)_l}, 
\label{eq_2}
\end{eqnarray}
with each component-specific edge probability vector ${\boldsymbol {\pi}}^{(h)}=({\pi}_1^{(h)}, \ldots, {\pi}_{V(V-1)/2}^{(h)})^\T \in (0,1)^{V(V-1)/2}$ defined as a function of a shared similarity vector ${\bf Z} \in \Re^{V(V-1)/2}$ and a component-specific one ${\bf D}^{(h)} \in \Re^{V(V-1)/2}$ characterized via matrix factorization representations. In particular, we let
\begin{eqnarray}
{\boldsymbol \pi}^{(h)}=\left[1+\exp\{{-{\bf Z}-{\bf D}^{(h)}}\}\right]^{-1},\quad {\bf D}^{(h)}={{\mathcal{L}}}({ X}^{(h)}{ \Lambda}^{(h)} { X}^{(h)\T}), \quad h=1, \ldots, H, 
\label{eq_3}
\end{eqnarray}
with the logistic mapping in \eqref{eq_3} applied element-wise. Equations  \eqref{eq_2}--\eqref{eq_3} carefully incorporate cluster dependence via cluster-specific mixing probabilities $\boldsymbol{\nu}_{k}=(\nu_{1k}, \ldots, \nu_{Hk})$, $k=1, \ldots, K$ as well as network information by considering a latent eigenmodel for each mixture component.

Focusing on the mixture component $h$, the latent eigenmodel \citep{hoff_2007} defines the undirected edges as realizations from conditionally independent Bernoulli random variables given their corresponding edge probabilities $\pi_l^{(h)}\in (0,1)$, $l=1, \ldots, V(V-1)/2$, and then borrows network information across these edge probabilities via lower dimensional representations. In particular, according to \eqref{eq_3} -- and letting $l$ correspond to the pair of products $v$ and $u$, $v>u$ -- each ${\pi}_l^{(h)}$ is constructed as a function of the pairwise similarity among products $v$ and $u$ in a latent space, with this similarity arising from the dot product of the products' latent coordinate vectors ${\bf X}_v^{(h)}=\{X_{v1}^{(h)}, \ldots, X_{vR}^{(h)} \}^{\T} \in \Re^{R}$, $v=1, \ldots, V$, with  ${\bf X}_v^{(h)\T}$ the $v$th row of ${X}^{(h)}$. Hence, products having coordinates in the same direction are more likely to be co-subscribed than products with coordinates in opposite directions, with the $R \times R$ matrix $\Lambda^{(h)}=\mbox{diag}(\boldsymbol{\lambda}^{(h)})=\mbox{diag}(\lambda^{(h)}_{1}, \ldots, \lambda^{(h)}_{R})$ weighting the similarity in each dimension $r$ by a non-negative parameter $\lambda^{(h)}_{r}$. Note that, consistently with our notation, the vector ${{\mathcal{L}}}({ X}^{(h)}{ \Lambda}^{(h)} { X}^{(h)\T})$ in \eqref{eq_3} is equal to {\small${({\bf X}_2^{(h)\T}{ \Lambda}^{(h)}{\bf X}_1^{(h)}, {\bf X}_3^{(h)\T}{ \Lambda}^{(h)}{\bf X}_1^{(h)}, \ldots, {\bf X}_V^{(h)\T}{ \Lambda}^{(h)}{\bf X}_1^{(h)},{\bf X}_3^{(h)\T}{ \Lambda}^{(h)}{\bf X}_2^{(h)},\ldots,{\bf X}_V^{(h)\T}{ \Lambda}^{(h)}{\bf X}_2^{(h)} ,\ldots,{\bf X}_V^{(h)\T}{ \Lambda}^{(h)}{\bf X}_{V-1}^{(h)} )^\T}$}.


The latent eigenmodel provides an appealing choice in reducing the dimensionality from $V(V-1)/2$ edge probabilities to $V \times R$ latent coordinates and $R$ weights -- typically $R\ll V$ -- and has been shown to provide a more flexible characterization of the connectivity patterns and network structures than stochastic block models \citep{now_2001}, latent distance models \citep{hof_2002} and mixed membership stochastic block models \citep{air_2008}. However, as discussed in \cite{dur_2015a}, a single latent eigenmodel fails in flexibly characterizing the probabilistic generative mechanism underlying a network-valued random variable. 

To improve flexibility and maintain computational tractability, equations   \eqref{eq_2}--\eqref{eq_3}  mix together $H$ latent eigenmodels, while adding a common similarity vector ${\bf Z} \in \Re^{V(V-1)/2}$ shared among all the co-subscription networks and centering the different mixture components to improve computational and clustering performance. According to \cite{dur_2015b}, this characterization guarantees full flexibility in approximating the cluster-specific  probability mass functions for the co-subscription networks.  Our goal is to exploit representation \eqref{eq_1}--\eqref{eq_3} in order to develop shared cross-sell strategies.

\begin{figure}[t]
\begin{center}
\includegraphics[width=0.9\textwidth]{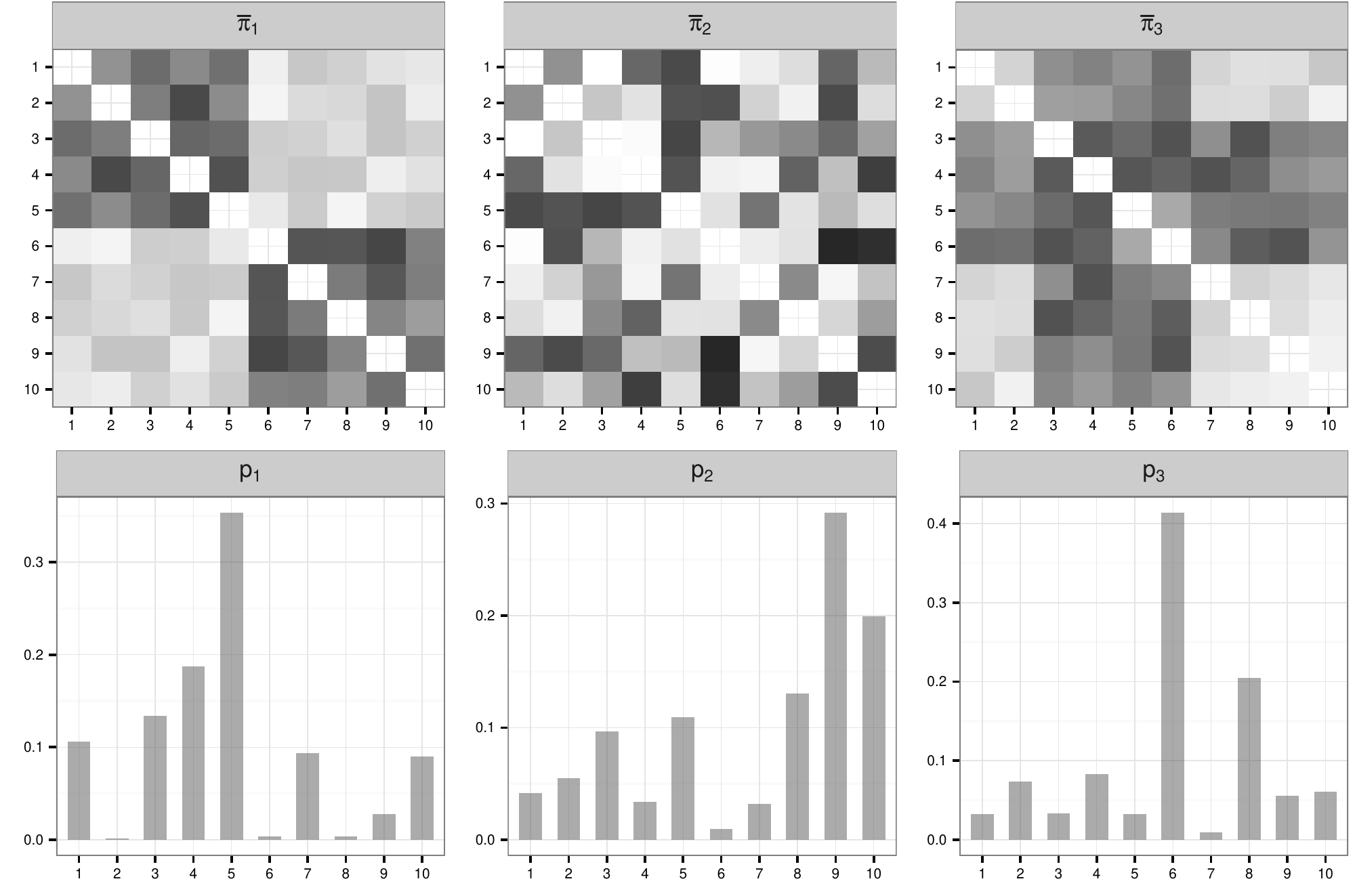}
\put (-355,250) {{{Cluster $1$}}}
\put (-230,250) {{{Cluster $2$}}}
\put (-110,250) {{{Cluster $3$}}}
\put (-340,-10) {{\small{Agencies in cluster $1$}}}
\put (-220,-10) {{\small{Agencies in cluster $2$}}}
\put (-97,-10) {{\small{Agencies in cluster $3$}}}
\put (-362,-35) {{\Huge{$\square  \ \ \ \ \square \ \ \ \ \square$}}}
\put (-355,-30) {{\footnotesize{$1$}}}
\put (-306,-30) {{\footnotesize{$4$}}}
\put (-257,-30) {{\footnotesize{$5$}}}
\put (-215,-35) {{\Huge{$\square \ \ \ \ \ \square$}}}
\put (-207,-30) {{\footnotesize{$2$}}}
\put (-150,-30) {{\footnotesize{$8$}}}
\put (-110,-35) {{\Huge{$\square \  \ \ \ \square \ \ \ \ \square$}}}
\put (-103,-30) {{\footnotesize{$3$}}}
\put (-53,-30) {{\footnotesize{$6$}}}
\put (-5,-30) {{\footnotesize{$7$}}}
\caption{\footnotesize{Example of a possible output from our model for cross-selling. Upper panels: cluster-specific edge probability vectors $\bar{\boldsymbol{\pi}}_1, \bar{\boldsymbol{\pi}}_2$ and $\bar{\boldsymbol{\pi}}_3$ -- rearranged in adjacency matrix form -- representing co-subscription profiles in each cluster. Color goes from white to black as the probability goes from $0$ to $1$. Lower panels: cluster-specific probability vectors ${\bf p}_1, {\bf p}_2$ and ${\bf p}_3$ representing mono-product customer choices in each cluster. Quantities in the upper panels are required to define the cross-sell strategies in each cluster. Quantities in the lower panels provide the additional information to define the associated performance indicators.}}
\label{F2}
\end{center}
\end{figure}
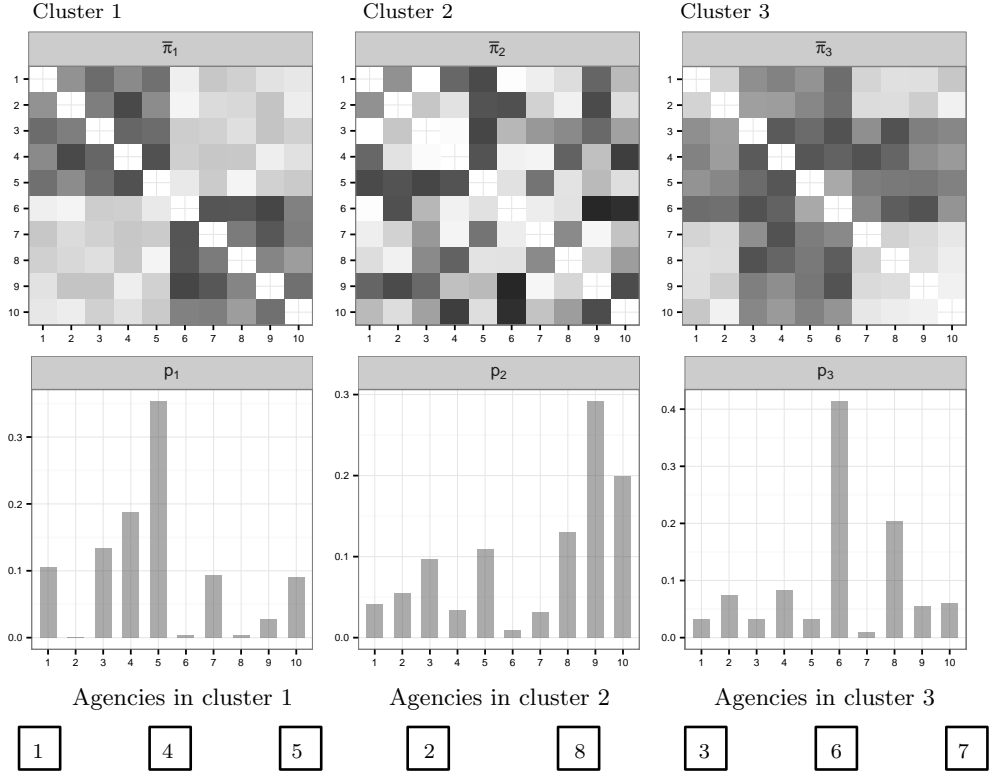

Figure \ref{F2} provides an example of the output from our model for  decision making in business intelligence when there are $n=8$ agencies and $K=3$ clusters. According to Figure  \ref{F2}, agencies $1$, $4$ and $5$ share common profiles of mono-product customer choices and co-subscription networks as $C_1=C_4=C_5=1$.  In Figure  \ref{F2},  mono-product choices in cluster $1$ are characterized by the vector ${\bf p}_{1}$, while the co-subscription behavior is summarized by the expectation of the network-valued random variable in cluster $1$, $\mbox{E}\{{\mathcal{L}}({\mathcal{A}}_1)\}=\bar{{\boldsymbol \pi}}_1=\sum_{{{\mathcal{L}(A)}} \in  \mathbb{A}_V} \mathcal{L}(A) \mbox{pr}\{{\mathcal{L}}({\mathcal{A}}_1)= \mathcal{L}(A)\}$, with $\mathbb{A}_V$ denoting the sample space of all the possible network configurations among $V$ products. As discussed in \cite{dur_2015b}, under representation \eqref{eq_2},  $\bar{{\boldsymbol \pi}}_1=\sum_{h=1}^H \nu_{h1} {\boldsymbol \pi}^{(h)}$, which coincides with the vector of co-subscription probabilities for pairs of products in cluster $1$. Hence, $\bar{{\boldsymbol \pi}}_1$  can be used to define the set of cross-sell strategies for agencies $1$, $4$ and $5$. The same description holds for clusters $2$ and $3$.

\begin{figure}[t]
\centering
\begin{tikzpicture}[scale=0.9, transform shape]
\tikzstyle{main}=[circle, minimum size = 13mm, thick, draw =black!80, node distance = 16mm]
\tikzstyle{connect}=[-latex, thick]
\tikzstyle{box}=[rectangle, draw=black!100]
  \node[main, fill = white!100] (theta) {$\boldsymbol{\nu}_k$ };
  \node[main] (z) [right=of theta] {$G_i$};
    \node[main,fill = black!10] (d) [below=of z] {${\bf y}_i$ };
        \node[main] (p) [left=of d] {${\bf p}_k$ };
  \node[main, fill = black!10] (w) [right=of z] {$\mathcal{L}({A}_i)$ };
    \node[main] (pi) [above=of w] {$\boldsymbol{\pi}^{(h)}$ };
        \node[main] (x) [right=of pi] {${X}^{(h)}$ };
                \node[main] (la) [left=of pi] {$\boldsymbol{\lambda}^{(h)}$ };
                                \node[main] (zeta) at (5.9,5) {${\bf Z}$ };
                                 \node[main] (c) at (4.5,-1.5) {$C_i$ };
     \path        (theta) edge [connect] (z)
        (z) edge [connect] (w)
                (c) edge [connect] (d)
                                (c) edge [connect] (z)
        (pi) edge [connect] (w)
                (p) edge [connect] (d)
                  (x) edge [connect] (pi)
                    (la) edge [connect] (pi)
                      (zeta) edge [connect] (pi);
  \node[rectangle, inner sep=6.5mm, draw=black!100, fit = (theta)] {};
  \node[rectangle, inner sep=0mm, fit= (theta)] {};
  \node[rectangle, inner sep=6.5mm, draw=black!100, fit = (p)] {};
    \node[rectangle, inner sep=5.5mm, draw=black!100, fit = (pi) (la) (x)] {};
  \node[] at (0.1,-1) {$k=1, \ldots, K$};
    \node[] at (0.1,-3.9) {$k=1, \ldots, K$};
        \node[] at (8.9,2) {$h=1, \ldots, H$};
\end{tikzpicture}
\caption{Graphical representation of the mechanism to generate $\{{\bf y}_i,\mathcal{L}(A_i)\}$, under representation \eqref{eq_1} and \eqref{eq_4}.}\label{F_3}
\end{figure}
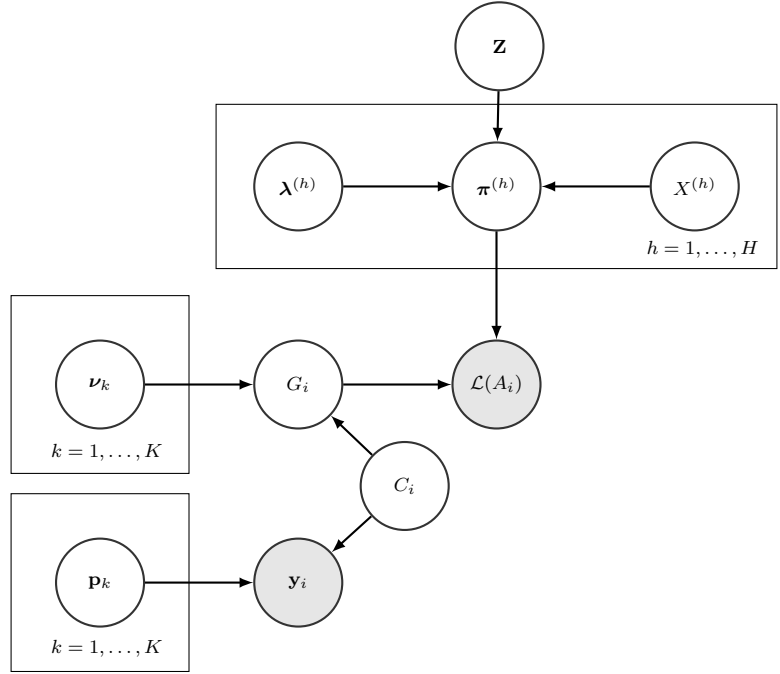

These quantities are estimated from our model and considered when defining the cluster-specific cross-sell strategies  $u_{k1}, \ldots, u_{kV}$ and computing their performance indicators  $e_{k1}, \ldots, e_{kV}$, for each $k=1, \ldots, K$. Adapting the initial discussion in Section \ref{sec_1} to the output of our model, cross-sell strategy $u_{kv}$ offers to mono-product customers who subscribed to $v$ in agencies within cluster $k$, the additional product $u_{kv}=\mbox{argmax}_u\{ \mbox{pr} (\mathcal{A}_{k[vu]}=1): u \neq v\}$. The probability of a co-subscription between products $v$ and $u$ for agencies in cluster $k$, $\mbox{pr} (\mathcal{A}_{k[vu]}=1)$, is easily available from our model as $\bar{{\pi}}_{kl}$, with $l$ the index denoting the pair $v$ and $u$ in the vectorized representation of the adjacency matrix. The performance indicator associated to $u_{kv}$ is instead $e_{kv}=p_{kv}\mbox{max}\{\mbox{pr} (\mathcal{A}_{k[vu]}=1): u \neq v\}$.

As motivated above, the main purpose of our analysis is to cluster agencies having customers with similar mono-product and multi-product purchasing behavior, while providing accurate estimates of the performance indicators for the different strategies.  In order to develop procedures for tractable estimation and inference, it is useful to rely on a hierarchical specification equivalent to \eqref{eq_2}--\eqref{eq_3}, which introduces an additional class index for each agency $i$, $G_i \in \{1, \ldots, H \}$, as follows: 
\begin{eqnarray}
\mathcal{L}(\mathcal{A}_i)_l \mid \pi_{il} & \stackrel{\mbox{{\tiny{indep}}}}{\sim}& \mbox{Bern}(\pi_{il}), \quad l=1, \ldots, V(V-1)/2,  \nonumber\\
{\boldsymbol \pi}_i \mid G_i=h, {\boldsymbol \pi}^{(h)}&=&{\boldsymbol \pi}^{(h)},  \quad {\boldsymbol \pi}^{(h)}=\left[1+\exp\{{-{\bf Z}-\mathcal{L}({X}^{(h)}{\Lambda}^{(h)} {X}^{(h)\T})}\}\right]^{-1}   \label{eq_4} \\
\mbox{pr}(G_i=h \mid C_i=k)&=&\nu_{hk}, \quad h=1, \ldots, H, \nonumber
\end{eqnarray}
independently for $i=1, \ldots, n$.  Recalling that $C_i$ is the cluster index for agency $i$, representation \eqref{eq_4} shows that agencies in the same cluster have a common set of probability weights over the components in the mixture model for the co-subscription network.  Moreover, although agencies in different clusters have different probabilities to be allocated in each mixture component, we are not forcing the class indices to be necessarily different for agencies belonging to different clusters. In fact, under  \eqref{eq_4}, two agencies $i$ and $j$ in different clusters $C_{i} \neq C_{j}$ are allowed to share the same component $G_{i}=G_{j}$ in the mixture model for the co-subscription networks. This facilitates efficient borrowing of information across clusters in modeling the mixture components characterizing the cluster-dependent mixture of latent eigenmodels. Refer to Figure \ref{F_3} for a graphical representation of our hierarchical model.

The next section develops a Bayesian approach for inference under the proposed model. Although model  \eqref{eq_2}--\eqref{eq_3} is reminiscent of recent proposals from \cite{dur_2015a} and  \cite{dur_2015b}, the procedure developed in this contribution has key differences.  \cite{dur_2015a} develop a nonparametric mixture of latent eigenmodels to characterize the probability mass function of a network-valued random variable.  \cite{dur_2015b} test for changes in this probability mass function across the values of a categorical predictor.  Our current focus is instead on using the  flexible and parsimonious joint model presented in Figure \ref{F_3} to develop shared cross-sell strategies.   In our business intelligence problem, the conditional distribution of the co-subscription networks varies as a function of a  latent clustering variable, which is endogenously determined by mono-product choices and co-subscription patterns shared across subsets of agencies.  These cluster assignments are key unknown quantities, requiring careful priors and adapted computational methods.

\section{Prior specification and posterior computation}\label{sec_3}

\subsection{Prior specification}
The assignment vector ${\bf C}=(C_1,\ldots,C_n)$ of the agencies and the total number of clusters $K$ are key unknowns in our analysis.  There is a rich literature on prior probability models for ${\bf C}$ and $K$.  A widely used approach is the Chinese Restaurant Process (CRP) \citep{ald_1985}, in which each cluster attracts new units in proportion to its size. In particular, letting  ${\bf C}=(C_1,\ldots,C_n) \sim \mbox{CRP}(\alpha_c)$, the prior distribution over clusters for the $i$th agency, conditioned on the membership of the others $C_1,\ldots ,C_{i-1},C_{i+1}, \ldots, C_{n}$, is
\begin{eqnarray}
\mbox{pr}(C_i=k \mid C_1,\ldots ,C_{i-1},C_{i+1}, \ldots, C_{n}) = \begin{cases} \frac{n_{k,-i}}{n-1+\alpha_c} & \quad \text{for} \ \ k=1, \ldots, K_{-i}, \\ \frac{\alpha_c}{n-1+\alpha_c} & \quad \text{for} \ \ k=K_{-i}+1,\\ \end{cases} 
 \label{CRP}
\end{eqnarray}
where $K_{-i}$ is the total number of nonempty clusters after removing $C_i$,  $n_{k,-i}$ is the total number of agencies allocated to cluster $k$, excluding the $i$th one, and $\alpha_c>0$ is a concentration parameter controlling the expected number of occupied clusters $\mbox{E}(K)= O( \alpha_c \log n)$. High values of $\alpha_c$ favor more clusters {\em a priori}, with $\mbox{E}(K)$ growing as sample size $n$ increases. 

Each cluster characterizes a specific joint profile of mono- and multi-product customer behaviors, with these profiles potentially arising from specific preferences of customers in different regions or economic areas where each agency operates. Differences in profiles may be also associated to particular business administration choices of each agency. Hence, we expect a growing $K$ as the number of agencies in the sample increases, with the sublinear relation between $\mbox{E}(K)$ and $n$ favoring parsimony in the number of cross-sell strategies to be defined. 

Equation \eqref{CRP} defines the conditional distribution of $C_i$ given the other assignments, after rearranging indices so that $1, \ldots, K_{-i}$ clusters are nonempty after removing $C_i$. These operations are possible since the cluster labels are arbitrary and observations are exchangeable based on the CRP prior; refer to \citet{griff_2011} and \citet{gers_2012} for an introductory overview. Although we focus on the CRP, our model can easily accommodate other commonly used priors for random partitions, such as the  Pitman-Yor process or the Kingman paintbox -- among others. Refer to \cite{hjo_2010} for a general overview. 

Accurate clustering of agencies also relies on careful modeling of the sequence of cluster-specific mono-product choices defined in \eqref{eq_1} and the collection of cluster-specific probabilistic generative mechanisms associated to the co-subscription networks via \eqref{eq_2}--\eqref{eq_3}. Efficient estimation of these quantities is fundamental to develop cross-sell strategies $u_{k1}, \ldots, u_{kV}$  in each cluster $k=1, \ldots, K$ and to quantify their performance via $e_{k1}, \ldots, e_{kV}$. Hence, we define large support priors, which do not rule out {\em a priori} any generative mechanism while maintaining tractable computations.
As ${\bf p}_{k}$ represents the probability mass function for the categorical random variable $\mathcal{Y}_k$, we let
\begin{eqnarray}
{\bf p}_{k}=(p_{k1}, \ldots, p_{kV}) \sim \mbox{Dirichlet}(\alpha_{1}, \ldots, \alpha_{V}), \quad k=1, \ldots, K.
 \label{mono}
\end{eqnarray}

The prior for the probabilistic generative mechanism of the co-subscription networks in each cluster is defined by choosing independent priors for the quantities in   \eqref{eq_2}--\eqref{eq_3}.  To maintain computational tractability, we consider independent Gaussian priors for the common similarities $Z_l  \sim \mbox{N}(\mu_l, \sigma_l^2)$, $l=1, \ldots, V(V-1)/2$ and standard Gaussians for the latent coordinates $X_{vr}^{(h)}\sim \mbox{N}(0,1)$, for each $v=1, \ldots, V$, $r=1, \ldots, R$ and $h=1, \ldots, H$. Adapting  \cite{rou_2011}, we choose independent Dirichlet priors with small hyperparameters for each cluster-specific  mixing probability vector ${\boldsymbol \nu}_{k}=(\nu_{1k}, \ldots, \nu_{Hk})\sim \mbox{Dirichlet}(1/H, \ldots, 1/H)$, $k=1, \ldots, K$ to favor deletion of redundant mixture components not required to characterize the co-subscription networks. Finally, as the dimension of the latent spaces in the eigenmodels is unknown, we consider a shrinkage prior on the weights vector ${\boldsymbol \lambda}^{(h)}$, $h=1, \ldots, H$ which  adaptively deletes redundant latent space dimensions not required to characterize the co-subscription probabilities. This is accomplished by choosing multiplicative inverse gamma $\mbox{MIG}(a_{1},a_{2})$ priors \citep{bha_2011}  for vectors ${\boldsymbol \lambda}^{(h)}$, $h=1, \ldots, H$, which favor shrinkage effects by inducing priors on elements $\lambda^{(h)}_{r}$ for $r=1, \ldots, R$, that are increasingly concentrated around 0 as $r$ increases, for appropriate $a_{2}$. Hence
\begin{eqnarray}
\lambda^{(h)}_{r} =\prod_{m=1}^{r}\frac{1}{\vartheta^{(h)}_{m}}, \quad \vartheta^{(h)}_{1}\sim \mbox{Ga}(a_1,1), \ \vartheta^{(h)}_{m>1}\sim \mbox{Ga}(a_2,1), \quad r=1, \ldots, R, \quad h=1, \ldots, H.
\label{eq_lam}
\end{eqnarray}  

Beside providing simple algorithms for posterior inference, adapting results in \cite{dur_2015b}, the above specification induces a prior on the probabilistic generative mechanism for the co-subscription networks in each cluster having full support. Full prior support  is  a  key  property  to  ensure  good  performance  in defining accurate cross-sell strategies,  because without prior support about the true data generating process,  the posterior cannot possibly  concentrate  around  the truth.

\subsection{Posterior computation}
Posterior computation is available via a simple Gibbs sampler outlined in Algorithms \ref{Algorithm_6} and \ref{Algorithm_7}, which exploits results in \citet{neal_2000} to allocate agencies to clusters under the CRP prior and steps in \cite{dur_2015a} to update the quantities in \eqref{eq_2}--\eqref{eq_3}.  \cite{dur_2015a}  join the standard data augmentation procedure for inference in mixture models outlined in \eqref{eq_4}, with a recent P\'olya-gamma data augmentation for conjugate inference in Bayesian logistic regression \citep{pol_2013}. 
\begin{breakablealgorithm}
\caption{Part I of the Gibbs sampler for joint modeling of mixed domain data}
\begin{algorithmic}
\STATE Conditionally on $(C_1, \ldots, C_n)$ update priors for quantities in equations \eqref{eq_1}, \eqref{eq_2} and \eqref{eq_3} as follows
\STATE --------------------------------------------------------------------------------------------------------------------------------
\vspace{-10pt}
\STATE {{\bf  [1] Update the cluster-specific mono-product choice probabilities in ${\bf p}_k$, $k=1, \ldots, K$}}
 \FOR{$k=1, \ldots, K$}
 \STATE Update ${\bf p}_k$ from $(p_{k1}, \ldots, p_{kV}) \mid -\sim \mbox{Dirichlet}\left(\alpha_1+ \sum_{i:C_i=k}n_{i1}, \ldots, \alpha_V+ \sum_{i:C_i=k}n_{iV} \right) $
\ENDFOR
\STATE --------------------------------------------------------------------------------------------------------------------------------
\vspace{-10pt}
\STATE  {{\bf [2] Allocate each co-subscription network to one of the mixture components}}
\FOR{$i=1, \ldots, n$}
\STATE Sample the class indicator $G_i$ from the discrete distribution with probabilities
\begin{eqnarray*}
\mbox{pr}(G_i=h \mid - )= \frac{ \nu_{hC_i} \prod_{l=1}^{V(V-1)/2}\{\pi_{l}^{(h)}\}^{ \mathcal{L}(A_i)_l} \{1-  \pi^{(h)}_{l}\}^{1- \mathcal{L}(A_i)_l}}{\sum_{q=1}^{H}\nu_{qC_i} \prod_{l=1}^{V(V-1)/2}\{\pi_{l}^{(q)}\}^{ \mathcal{L}(A_i)_l} \{1-  \pi^{(q)}_{l}\}^{1- \mathcal{L}(A_i)_l}},
\end{eqnarray*}
for each $h=1, \ldots, H$, with ${\boldsymbol \pi}^{(h)}$ defined in \eqref{eq_3}.
\ENDFOR
\STATE --------------------------------------------------------------------------------------------------------------------------------
\vspace{-10pt}
\STATE  {{\bf [3] Update the cluster-specific vectors of mixing probabilities ${\boldsymbol \nu}_k$, $k=1, \ldots, K$}}
\FOR{$k=1, \ldots, K$}
\STATE Update each ${\boldsymbol \nu}_{k}$ from $(\nu_{1k}, \ldots, \nu_{Hk}) \mid -\sim \mbox{Dirichlet}(1/H+n_{1k}, \ldots, 1/H+n_{Hk})$ with $n_{hk}$ \\ denoting the number of agencies in cluster $k$, currently associated to mixture component $h$.
\ENDFOR
\STATE --------------------------------------------------------------------------------------------------------------------------------
\vspace{-10pt}
\STATE  {{\bf [4] Update quantities ${\bf Z}$, ${X}^{(h)}$ and ${\boldsymbol \lambda}^{(h)}$, $h=1, \ldots, H$}} via Poly\'a-gamma data augmentation exploiting derivations in \cite{dur_2015a} as follows
\FOR{$h=1, \ldots, H$}
\STATE{[4.1]} For each  $l=1, \ldots, V(V-1)/2$ update augmented data $\omega^{(h)}_{l}$ from the P\'olya-gamma
\begin{eqnarray*}
\omega^{(h)}_{l} \mid - \sim \mbox{\small{PG}}\left\{n_h, Z_l+\mathcal{L}(X^{(h)}\Lambda^{(h)} X^{(h)\T})_l\right\},
\end{eqnarray*}
with  $\mbox{\small{PG}}(b,c)$ denoting the  P\'olya-gamma distribution with parameters $b>0$ and $c \in \Re$, and $n_h$ the total number of agencies in mixture component $h$.
\STATE -----------------------------------------------------------------------------------------------------------------------------
\vspace{-10pt}
\STATE{[4.2]} Update the shared similarity vector ${\bf Z}$ from 
\begin{eqnarray*}
{\bf Z} \mid - \sim \mbox{N}_{V(V-1)/2}\{{\boldsymbol \mu}_Z, \mbox{diag}(\sigma^2_{Z_1}, \ldots, \sigma^2_{Z_{V(V-1)/2}})\},
\end{eqnarray*}
with ${\boldsymbol \mu}_Z$ having elements $\mu_{Z_l}=\sigma^2_{Z_l}[\sigma_l^{-2}\mu_l+\sum_{h=1}^{H}\{\mathcal{L}(A^{(h)})_l -n_h/2-\omega^{(h)}_{l}\mathcal{L}(X^{(h)} \Lambda^{(h)} X^{(h)\T})_l\}]$, where $\sigma^2_{Z_l}=1/(\sigma^{-2}_l+\sum_{h=1}^{H}\omega^{(h)}_{l})$, for each $l=1, \ldots, V(V-1)/2$ and $\mathcal{L}(A^{(h)})_l=\sum_{i:G_i=h}\mathcal{L}(A_i)_l$.
\STATE -----------------------------------------------------------------------------------------------------------------------------
\vspace{-10pt}
\STATE{[4.3]} As ${\bf D}^{(h)}={{\mathcal{L}}}({ X}^{(h)}{ \Lambda}^{(h)} { X}^{(h)\T})=\mathcal{L}({\bar{X}}^{(h)}{\bar{X}}^{(h)\T})$, with  ${\bar{X}}^{(h)}={X}^{(h)}{\Lambda}^{{(h)}1/2}$, update  elements in ${\bar{X}}^{(h)}$, knowing that, under our prior specification $\bar{X}_{vr}^{(h)}| \lambda_r^{(h)} \sim \mbox{N}(0, \lambda^{(h)}_r)$. Therefore
\FOR{$v=1, \ldots, V$}
\STATE Block-sample the $v$th row of ${\bar{X}}^{(h)}$. 
Define $ \bar{\bf X}^{(h)}_v=(\bar{X}^{(h)}_{v1}, \ldots, \bar{X}^{(h)}_{vR})^\T$ and let ${\bar{X}}^{(h)}_{(-v)}$ denote the $(V-1) \times R$ matrix obtained by removing the $v$th row in ${\bar{X}}^{(h)}$. Consider the logistic regression 
\begin{eqnarray*}
\mathcal{L}(A^{(h)})_{(v)}  \sim \mbox{Binom}(n_h,{\boldsymbol  \pi}_{(v)}^{(h)}), \quad \mbox{logit}({\boldsymbol  \pi}_{(v)}^{(h)}) ={\bf Z}_{(v)}+{\bar{X}}^{(h)}_{(-v)}{\bar{\bf X}}^{(h)}_v,
\label{eq_logis}
\end{eqnarray*}
with $\mathcal{L}(A^{(h)})_{(v)}$ and ${\bf Z}_{(v)}$ obtained by stacking elements $\mathcal{L}(A^{(h)})_l$ and $Z_l$, respectively, for all $l$ corresponding to pairs having $v$ as a one of the two nodes, and ordered consistently with the linear predictor. Exploiting previous logistic regression, and letting ${\Omega}_{(v)}^{(h)}$ be the diagonal matrix with the corresponding P\'olya-gamma augmented data, the full conditional is
{\small{
\begin{eqnarray*}
{\bar{\bf X}}^{(h)}_v \mid - \sim \mbox{N}_{R}\left\{\left({\bar{X}}^{(h)\T}_{(-v)}{\Omega}_{(v)}^{(h)} {\bar{X}}^{(h)}_{(-v)}+{{\Lambda}^{(h)}}^{-1}\right)^{-1} {\boldsymbol \eta}^{(h)}_{v}, \left({\bar{X}}^{(h)\T}_{(-v)}{\Omega}_{(v)}^{(h)} {\bar{X}}^{(h)}_{(-v)}+{{\Lambda}^{(h)}}^{-1}\right)^{-1}\right\},
\end{eqnarray*}}}
with $ {\boldsymbol  \eta}^{(h)}_{v}={\bar{X}}^{(h)\T}_{(-v)}\{\mathcal{L}(A^{(h)})_{(v)}-{1}_{V-1} n_h/2-{\Omega}_{(v)}^{(h)}{\bf Z}_{(v)}\}$
\ENDFOR
\STATE -----------------------------------------------------------------------------------------------------------------------------
\vspace{-10pt}
\STATE {[4.4] Sample the gamma quantities defining the shrinkage weights $\lambda_1^{(h)}, \ldots, \lambda_R^{(h)}$}
\vspace{-10pt}
\STATE \begin{eqnarray*}
\vartheta_{1}^{(h)} \mid - &\sim& \mbox{Ga} \left\{a_{1}+\frac{V R}{2},1+\frac{1}{2}\sum_{m=1}^{R}\theta_{m}^{(-1)}\sum_{v=1}^{V}(\bar{X}_{vm}^{(h)})^{2}\right\}, \quad \quad \quad \quad \nonumber \\
\vartheta^{(h)}_{r>1} \mid - &\sim& \mbox{Ga}\left\{a_{2}+\frac{V(R-r+1)}{2},1+\frac{1}{2}\sum_{m=r}^{R}\theta_{m}^{(-r)}\sum_{v=1}^{V} (\bar{X}_{vm}^{(h)})^{2}\right\},
\end{eqnarray*}
where $\theta_{m}^{(-r)}=\prod_{t=1,t\neq r}^{m} \vartheta^{(h)}_{t}$ for $r=1,\ldots,R$.
\ENDFOR
\STATE --------------------------------------------------------------------------------------------------------------------------------
\vspace{-10pt}
\STATE  {{\bf [5] Update the edge probability vectors ${\boldsymbol \pi}^{(h)}$, in each mixture component $h=1, \ldots, H$}}
\FOR{$h=1, \ldots, H$}
\STATE Compute ${\boldsymbol \pi}^{(h)}$ as ${\boldsymbol \pi}^{(h)}=\left[1+\exp\{{-{\bf Z}-\mathcal{L}(\bar{X}^{(h)} \bar{X}^{(h)\T})}\}\right]^{-1}$ 
\ENDFOR

\end{algorithmic}
\label{Algorithm_6}
\end{breakablealgorithm}

Algorithm \ref{Algorithm_6} provides detailed steps to update the cluster-specific probabilistic representation  of the mono-product customer choices and the cluster-specific probabilistic generative mechanism underlying co-subscription networks, given the cluster assignments $C_1, \ldots, C_n$. 
In performing these steps, the number of mixture components $H$ and the dimension of the latent spaces $R$ are set at conservative upper bounds, allowing the shrinkage priors for these quantities to adaptively empty redundant components that are not required to characterize the observed data. If all the mixture components are occupied or the posteriors for some weights $\lambda^{(h)}_{r}$ are not concentrated around $0$ for any $h$, this suggests that $H$ or $R$ should be increased, respectively.

All the previous steps are straightforward to compute, exploiting the data augmentation strategy described in \eqref{eq_4} for updating quantities in \eqref{eq_2}--\eqref{eq_3}. The latter provides key computational benefits also when sampling from the full conditional of the cluster assignments described in Algorithm \ref{Algorithm_7}.

\begin{breakablealgorithm}
\caption{Part II of the Gibbs sampler for joint modeling of mixed domain data}
\begin{algorithmic}
\STATE Conditionally on samples for the quantities   in \eqref{eq_1}, \eqref{eq_2} and \eqref{eq_3}, update the cluster indices in ${\bf C}$.
\STATE --------------------------------------------------------------------------------------------------------------------------------
\vspace{-10pt}
\STATE  {{\bf [6] Sample the cluster assignments $(C_1, \ldots, C_n)$ via a sequential re-seating procedure}}
\FOR{$i=1, \ldots, n$}
\STATE Update $C_{i}$ conditionally on $C_{-i}=(C_1,\ldots ,C_{i-1},C_{i+1}, \ldots, C_{n})$ 
\STATE [6.1] Remove agency $i$ since we are going to sample its cluster membership $C_i$.
\STATE [6.2] If no other agencies are in the same cluster of $i$, this cluster becomes empty and is removed along with its associated mono-product choice probabilities and co-subscription network probability mass function.
\STATE [6.3] Re-order cluster indices so that $1, \ldots, K_{-i}$ are nonempty.
\STATE [6.4] Update the cluster of $i$ from the full conditional categorical variable with cluster probabilities\begin{flalign}
&\mbox{pr}(C_i=k \mid - )\propto \begin{cases} \frac{n_{k,-i}}{n-1+\alpha_c}\mbox{pr}\left\{\mathcal{L}(A_i),{\bf y}_i, G_i \mid C_i=k,  {\boldsymbol \pi}^{(G_i)}, {\bf p}_{k}, {\boldsymbol \nu}_k \right\} &  \text{for} \ \ k=1,\ldots, K_{-i}, \\ \frac{\alpha_c}{n-1+\alpha_c}\mbox{pr}\left\{\mathcal{L}(A_i), {\bf y}_i, G_i \mid C_i=K_{-i}+1, {\boldsymbol \pi}^{(G_i)}\right\} &  \text{for} \ \  k=K_{-i}+1,
\end{cases} 
\label{color_code}&
\end{flalign}
\STATE [6.5] If $i$ is assigned a new cluster $K_{-i}+1$, add a new cluster and sample a new ${\bf p}_{K_{-i}+1}$ and ${\boldsymbol \nu}_{K_{-i}+1}$ conditionally on ${\bf y}_i$ and $G_i$ according to steps {\bf{[1]}} and {\bf{[3]}}, respectively.
\ENDFOR
\end{algorithmic}
\label{Algorithm_7}
\end{breakablealgorithm}
To perform steps in Algorithm \ref{Algorithm_7}  one needs to compute the conditional probabilities in  equation \eqref{color_code} at each MCMC iteration. Although this is apparently a cumbersome task, our model formulation \eqref{eq_1}--\eqref{eq_3} along with its hierarchical  representation in \eqref{eq_4} allows key simplifications, substantially improving computational tractability. Specifically, under our model, the conditional probability  $\mbox{pr}\left\{\mathcal{L}(A_i), {\bf y}_i, G_i \mid C_i=k,   {\boldsymbol \pi}^{(G_i)}, {\bf p}_{k}, {\boldsymbol \nu}_k\right\} $ for augmented data -- including $G_i$ -- can be factorized as
{\small{
\begin{eqnarray}
\mbox{pr}\{\mathcal{L}(A_i) \mid G_i, {\boldsymbol\pi}^{(G_i)}\}\mbox{pr}({\bf y}_i \mid C_i=k,  {\bf p}_{k})\mbox{pr}(G_i \mid C_i=k,  {\boldsymbol \nu}_k)=\mbox{pr}\{\mathcal{L}(A_i) \mid G_i, {\boldsymbol\pi}^{(G_i)}\} \prod_{v=1}^V p_{kv}^{n_{iv}} \prod_{h=1}^H \nu_{hk}^{1_{(h)}(G_i)}
\label{cond_lik}
\end{eqnarray}}}with $1_{(h)}(G_i)=1$ if $G_i=h$ and $1_{(h)}(G_i)=0$, otherwise. According to  \eqref{cond_lik} inducing cluster-dependence through the mixing probabilities ${\boldsymbol \nu}_k$, while considering cluster-independent mixture components in \eqref{eq_2}--\eqref{eq_3}, has the key benefit of  maintaining $\mbox{pr}\{\mathcal{L}(A_i) \mid G_i,  {\boldsymbol\pi}^{(G_i)}\}$ constant across the cluster assignments. Recalling formulation \eqref{eq_4}, conditionally on  $G_i=h$, the probability of observing $\mathcal{L}(A_i)$ is 
\begin{eqnarray*}
\mbox{pr}\{\mathcal{L}(A_i) \mid G_i=h, {\boldsymbol\pi}^{(h)}\}=\prod_{l=1}^{V(V-1)/2} \left\{\pi_{l}^{(h)}\right\}^{\mathcal{L}({{A}}_i)_l} \left\{1-  \pi^{(h)}_{l}\right\}^{1-\mathcal{L}({{A}}_i)_l},
\end{eqnarray*}
and  does not depend on the cluster assignment $C_i$. As a result $\mbox{pr}\{\mathcal{L}(A_i) \mid G_i, {\boldsymbol\pi}^{(G_i)}\}$ acts as a multiplicative constant in \eqref{color_code}, allowing  \eqref{color_code} to be further simplified as
\begin{eqnarray}
&\mbox{pr}(C_i=k \mid - )\propto \begin{cases} \frac{n_{k,-i}}{n-1+\alpha_c}  \prod_{v=1}^V p_{kv}^{n_{iv}} \prod_{h=1}^H \nu_{hk}^{1_{(h)}(G_i)}&    \text{for} \  k=1, \ldots, K_{-i}, \\ \frac{\alpha_c}{n-1+\alpha_c}\mbox{pr}({\bf y}_i \mid C_i=K_{-i}+1)\mbox{pr}(G_i \mid C_i=K_{-i}+1) &    \text{for} \   k=K_{-i}+1,
\end{cases} 
\label{color_code_1}
\end{eqnarray}
where the two marginal probabilities corresponding to a newly occupied cluster are easily available exploiting the multinomial-Dirichlet conjugacy. In particular, it is easy to show that 
\begin{eqnarray}
\mbox{pr}({\bf y}_i \mid C_i=K_{-i}+1)= \int  \prod_{v=1}^V p_{K_{-i}+1, v}^{n_{iv}} d \Pi({\bf p}_{K_{-i}+1})=\frac{\Gamma(\sum_{v=1}^V \alpha_v)}{\prod_{v=1}^V \Gamma(\alpha_v)}\frac{\prod_{v=1}^V \Gamma(\alpha_v+n_{iv})}{\Gamma\{\sum_{v=1}^V (\alpha_v+n_{iv})\}},
\label{integrated_mono}
\end{eqnarray}
for the mono-product choices, and
\begin{eqnarray}
\mbox{pr}(G_i \mid C_i=K_{-i}+1)= \int  \prod_{h=1}^H \nu_{h, K_{-i}+1}^{1_{(h)}(G_i)} d \Pi({\boldsymbol \nu}_{K_{-i}+1})=\frac{\Gamma(\sum_{h=1}^H 1/H)}{\prod_{h=1}^H \Gamma(1/H)}\frac{\prod_{h=1}^H \Gamma\{1/H+1_{(h)}(G_i)\}}{\Gamma[\sum_{h=1}^H \{1/H+1_{(h)}(G_i)\}]},
\label{integrated_multi}
\end{eqnarray}
for the class indicator variable in the cluster-dependent mixture of latent eigenmodels.  

Hence, considering only cluster dependence in the mixing probability vectors ${\boldsymbol \nu}_k$, $k=1, \ldots, K$ and exploiting the augmented data $G_i$, $i=1, \ldots, n$, in the mixture representation for the generative mechanism underlying co-subscription networks, allows a massive gain in computational tractability for step {\bf{[6]}} in Algorithm  \ref{Algorithm_7}. While \eqref{integrated_mono} and \eqref{integrated_multi} can be easily derived in closed form, the marginal probability of the co-subscription networks with respect to the edge probability vectors arising from the construction in \eqref{eq_4} is not analytically available.

\section{Simulation studies}
\label{sec_4}
We consider a simulation study to evaluate the performance of our method in accurately recovering clusters of agencies and in efficiently estimating the key quantities required to define the set of cross-sell strategies for each cluster and their associated performance indicators. In simulating data, we define a scenario mimicking the structure of our application.

We focus on $n=200$ agencies equally divided in $K=4$ latent clusters and consider a total number of $V=15$ products. Graphical analyses of our data -- highlighted in Figure \ref{F1} --  show that mono-product customers typically concentrate on a small subset of the available products with high probability, while choosing the remaining set with very low frequency. We maintain this behavior in constructing ${\bf p}^0_{k}$, $k=1, \ldots, 4$, while defining a challenging scenario with small changes in ${\bf p}^0_{k}$ across clusters.  In particular, we set $p^{0}_{1v}$ and $p^{0}_{2v}$ to be equal for all products $v$ except for permuting $1$ and $9$, so $\mbox{pr}(\mathcal{Y}^0_1=1)=\mbox{pr}(\mathcal{Y}^0_2=9)$ and  $\mbox{pr}(\mathcal{Y}^0_1=9)=\mbox{pr}(\mathcal{Y}^0_2=1)$. We adopt a similar strategy for clusters 3 and 4 by considering $p^{0}_{3v}$ and $p^{0}_{4v}$ equal for all products $v$ except $3$ and $7$, where we let  $\mbox{pr}(\mathcal{Y}^0_3=3)=\mbox{pr}(\mathcal{Y}^0_4=7)$ and $\mbox{pr}(\mathcal{Y}^0_3=7)=\mbox{pr}(\mathcal{Y}^0_4=3)$. We simulate mono-product subscription data $y_{is}$, $i=1, \ldots, 200$ and $s=1, \ldots, 500$, from the categorical random variable $\mathcal{Y}^0_k$ with probability mass function ${\bf p}^0_{k}$, where $k=1$ for agencies $i=1, \ldots, 50$, $k=2$ for $i=51, \ldots, 100$, $k=3$ for $i=101, \ldots, 150$ and $k=4$ for $i=151, \ldots, 200$. Although agencies in our application have at least $\approx 1{,}000$ mono-product customers, we consider a smaller number $n_i=500$ for each agency $i=1, \ldots, 200$ to evaluate the performance when there is less data.

Co-subscription networks are simulated exploiting our constructive representation of the mixture model in \eqref{eq_2}. We consider $H=3$ mixture components with each edge probability vector ${\boldsymbol \pi}^{0(h)}$ generated to mimic a possible co-subscription scenario. Specifically, ${\boldsymbol \pi}^{0(1)}$ is characterized by one dense community among 10 possibly highly related products, while assigning low probability to the remaining pairs of products.  Vector ${\boldsymbol \pi}^{0(2)}$ represents the case of 4 hub products, which occur with high probability in multi-product customer choices, while fixing the remaining co-subscription probabilities at low values. To reduce separation among mixture components and provide a more challenging scenario, the edge probability vector  ${\boldsymbol \pi}^{0(3)}$ is very similar to ${\boldsymbol \pi}^{0(2)}$ with exception of product $v=4$ which is held out from the hub products. In avoiding the eigenmodel construction \eqref{eq_3} in the definition of ${\boldsymbol \pi}^{0(h)}$, $h=1, \ldots, 3$, we additionally aim to evaluate the performance of representation \eqref{eq_3} in accurately characterizing the co-subscription probabilities for each component $h$.

In simulating vectors $\mathcal{L}(A_i)$, $i=1, \ldots, 200$, we consider cluster-specific mixing probability vectors ${\boldsymbol\nu}^0_1={\boldsymbol\nu}^0_2=(0.9,0.05,0.05)$, ${\boldsymbol\nu}^0_3=(0.05,0.9,0.05)$ and ${\boldsymbol \nu}^0_4=(0.05,0.05,0.9)$. This choice allows the first co-subscription scenario defined by ${\boldsymbol \pi}^{0(1)}$ to be very likely in agencies belonging to clusters $1$ and $2$. Scenarios characterized by ${\boldsymbol\pi}^{0(2)}$ and ${\boldsymbol \pi}^{0(3)}$ are instead more likely in clusters $3$ and $4$, respectively. Letting ${\boldsymbol\nu}^0_1={\boldsymbol\nu}^0_2$ further reduces separation among clusters $1$ and $2$. These two clusters have very similar mono-product choice probabilities  and equal generative processes for the co-subscription networks, providing a challenging scenario to evaluate clustering performance.

We analyze the simulated data under our model  \eqref{eq_1}--\eqref{eq_3}. As in \cite{dur_2015a}, we set $a_{1}=2.5$, $a_{2}=3.5$ and  $\sigma^2_{l}=10$, $l=1, \ldots, V(V-1)/2$. Quantities $\mu_{l}$, $l=1, \ldots, V(V-1)/2$ are defined as $\mu_{l}=\mbox{logit}\{\sum_{i=1}^{n} \mathcal{L}(A_{i})_l/n\}$ in order to center the mixture representation for the co-subscription networks around a structure shared by all the agencies in the company. We adopt a similar strategy for the hyperparameters of the Dirichlet prior \eqref{mono} by setting $\alpha_{v}=\sum_{i=1}^{n} n_{iv}/n$ for each $v=1, \ldots, V$, in order to center  \eqref{mono} around the averaged preferences of mono-product customers in the entire company. This empirical choice improves performance relative to symmetric Dirichlet priors, which can be sensitive to the concentration parameter, overly penalizing the addition of new clusters when concentration is small.  Finally, we set the concentration parameter $\alpha_c=1$ in the CRP prior for the cluster assignments,  according to standard practice. Our approach can be easily generalized to learn $\alpha_c$ from the data as in \citet{esco_1995}.  However, we obtain the same results in term of clustering, cross-sell strategies and performance indicators when using $\alpha_c=0.5$, $\alpha_c=5$, $\alpha_c=10$, $\alpha_c=15$ and $\alpha_c=20$.

\begin{figure}
\centering
\includegraphics[height=7.5cm, width=14.5cm]{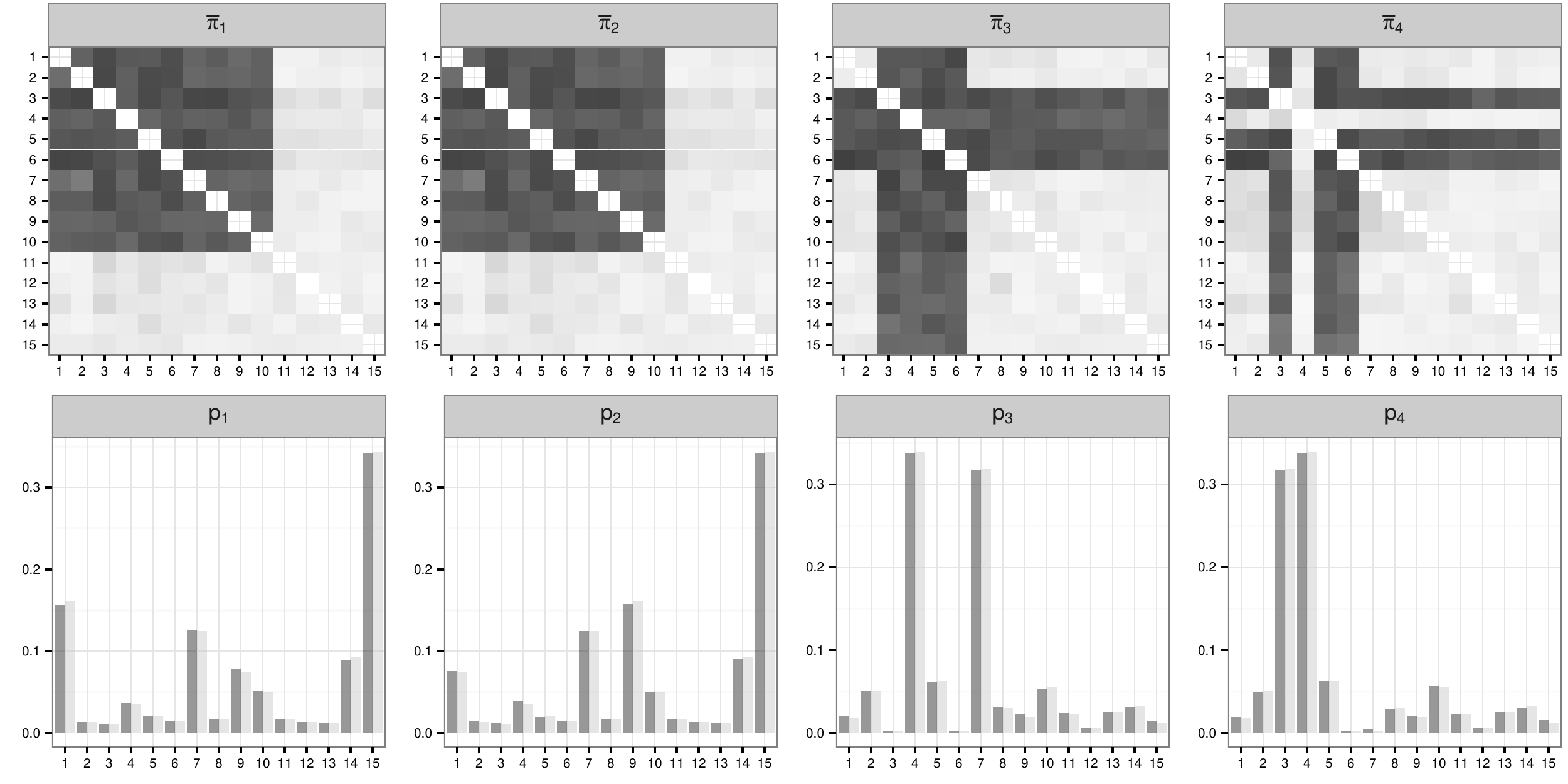}
\caption{\footnotesize{For the four nonempty clusters. Upper panels: posterior mean $\hat{\bar{\boldsymbol \pi}}_k$ of the co-subscription probabilities among pairs of products in each cluster $k=1, \ldots, 4$ (lower triangular) and true ${\bar{\boldsymbol \pi}}^{0}_k$, $k=1, \ldots, 4$ (upper triangular). Color goes from white to black as the probability goes from $0$ to $1$. Lower panels: posterior mean $\hat{\bf p}_{k}$ of the mono-product choices in each cluster $k=1, \ldots, 4$ (dark gray) and true ${\bf p}^0_{k}$, $k=1, \ldots, 4$ (light gray).}}
\label{F4}
\end{figure}

We collect $5{,}000$ Gibbs iterations and set $H=15$ and $R=10$ as upper bounds for the number of mixture components and the dimension of the latent spaces, respectively. These upper bounds provide a good choice, with the sparse Dirichlet prior for the mixing probabilities in ${\boldsymbol \nu}_k$, $k=1, \ldots, K$, and the multiplicative inverse gamma for the weights in ${\boldsymbol \lambda}^{(h)}$, $h=1, \ldots, H$, adaptively removing redundant components.  Trace plots suggest convergence is reached after a burn-in of $1{,}000$ and mixing for the quantities of interest for inference is good. As inference focuses on cluster-specific structures, it is important to first check for label switching issues, and relabel the clusters at each MCMC iteration using, for example, \cite{ste_2000} as needed.
Trace plots suggest label switching is not an issue in our simulation.

We initialize our MCMC algorithm by assigning all agencies to a single cluster.  Then, using samples after burn-in, we estimate $\hat{\bf C}$ as corresponding to the partition of agencies observed in the maximum proportion of draws, providing an estimate of the MAP.  We found this simple approach to have good performance and stability in our simulations, but as the number of agencies increase using more refined procedures, such as  \cite{med_2004} and \cite{lau_2007}, may be preferable.  We estimate $\hat{K}=4$, correctly grouping all the simulated agencies, including those in clusters $1$ and $2$, which are characterized by very subtle differences in their generating process.

\begin{figure}
\centering
\includegraphics[height=5cm, width=15cm]{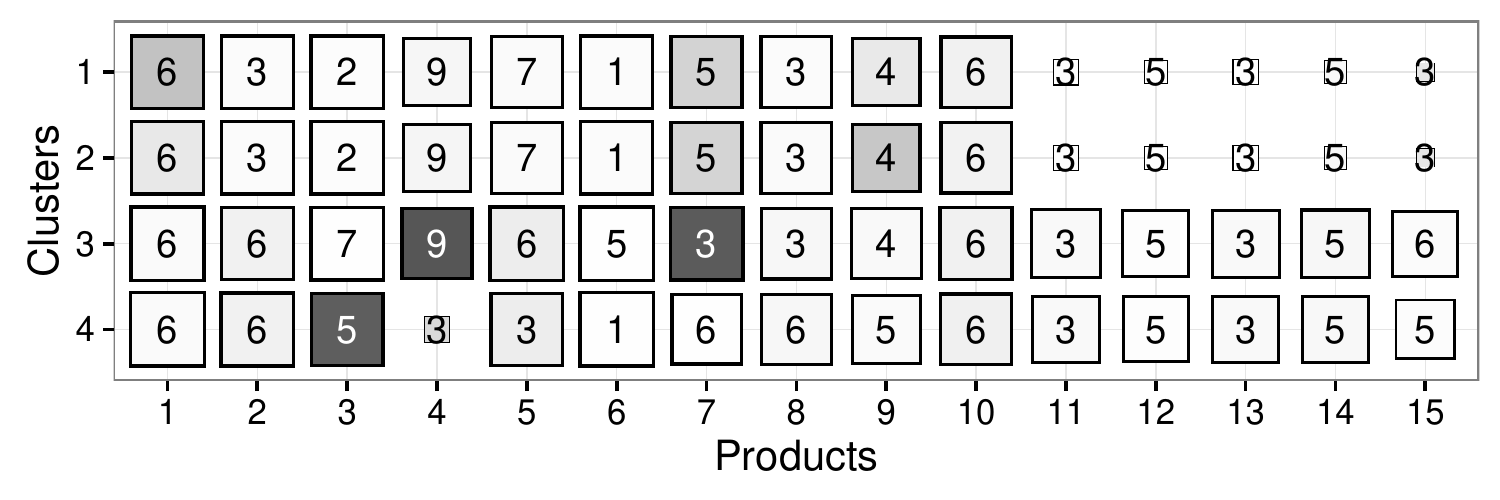}
\caption{\footnotesize{Summary of the estimated cluster-specific cross-sell strategies $\hat{u}_{k1}, \ldots, \hat{u}_{kV}$ along with their performance indicators  $\hat{e}_{k1}, \ldots, \hat{e}_{kV}$. Each cell $[k,v]$ of the matrix defines the cross-sell strategy for mono-product customers subscribed to product $v$ in agencies belonging to cluster $k$, along with its performance. The number in cell $[k,v]$ corresponds to the best offer $ \hat{u}_{kv}=\mbox{argmax}_u\{ \hat{\mbox{pr}}(\mathcal{A}_{k[vu]}=1): u \neq v\}$. The dimension of each square is proportional to $\mbox{max}\{ \hat{\mbox{pr}}(\mathcal{A}_{k[vu]}=1): u \neq v\}$, while the gray scale of the color is proportional to the corresponding estimated performance $\hat{e}_{kv}=\hat{p}_{kv}\mbox{max}\{ \hat{\mbox{pr}}(\mathcal{A}_{k[vu]}=1): u \neq v\}$. The estimated probability of a co-subscription for each pair of products $v$ and $u$ in agencies within cluster $k$, $\hat{\mbox{pr}}(\mathcal{A}_{k[vu]}=1)$, is easily available from the posterior mean $\hat{\bar{{\pi}}}_{kl}$ of ${\bar{{\pi}}}_{kl}$ for each $k=1, \ldots, K$ and $l=1, \ldots, V(V-1)/2$. 
}}
\label{F5}
\end{figure}

Accurate clustering further allows for efficient estimation of the cluster-specific components required for cross-selling. According to Figure \ref{F4}, we correctly estimate the probability mass functions ${\bf p}^0_{k}$ characterizing the mono-product choices in each cluster, with a similar performance in recovering the vectors of co-subscription probabilities $\bar{\boldsymbol\pi}^0_k$.  These estimates are a key to define the cluster-specific cross-sell strategies $u_{k1}, \ldots, u_{kV}$ and related performance indicators $e_{k1}, \ldots, e_{kV}$, as shown in Figure \ref{F5}. Consistently with results in Figure  \ref{F4}, cross-sell strategies are the same in clusters $1$ and $2$ as $\hat{\bar{\boldsymbol \pi}}_{1} \approx \hat{ \bar{\boldsymbol \pi}}_{2}$, while performance indicators differ only for strategies targeting mono-product customers subscribed to $v=1$ or $v=9$. The first segment is more profitable in cluster $1$ while the second is in cluster $2$. This is consistent with our estimates in Figure  \ref{F4} highlighting $\hat{\mbox{pr}}(\mathcal{Y}_1=1)>\hat{\mbox{pr}}(\mathcal{Y}_1=9)$ and $\hat{\mbox{pr}}(\mathcal{Y}_2=9)>\hat{\mbox{pr}}(\mathcal{Y}_2=1)$. Mono-product customers subscribed to $v=4$ and $v=7$ are highly profitable in cluster $3$ in being highly represented and having high co-subscription probability with at least one further product.  Although $v=4$ is highly populated in cluster $4$, according to $\hat{\bar{\boldsymbol  \pi}}_{4}$ it is not possible to find another product $u$ having high co-subscription probability with $v=4$, so such customers are not profitable in terms of targeting of cross-sell campaigns.

\begin{figure}
\centering
\includegraphics[height=5.3cm, width=13cm]{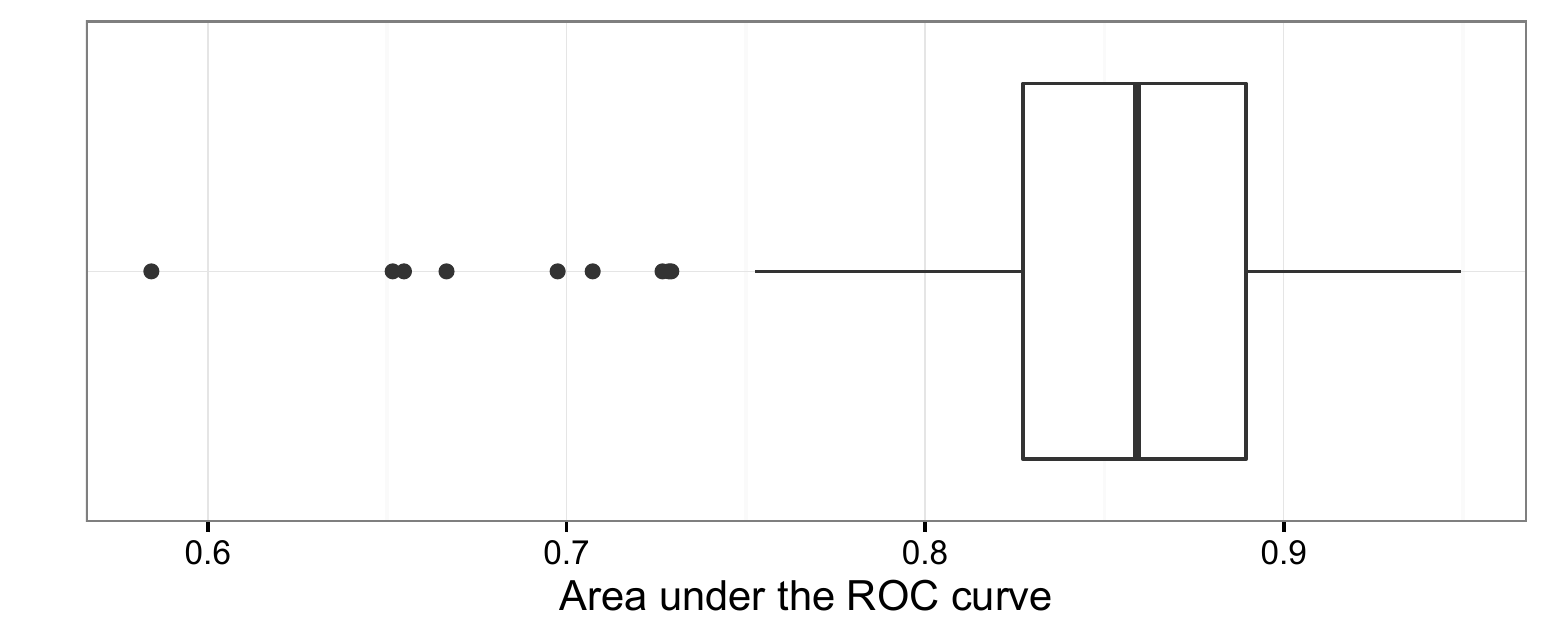}
\caption{\footnotesize{Box plot of the areas under the ROC curve $\mbox{AUC}_1, \ldots, \mbox{AUC}_{200}$. Each $\mbox{AUC}_i$ is  obtained using the simulated  $\mathcal{L}(A_{i})$ and the estimated co-subscription probability vector $\hat{\bar{\boldsymbol  \pi}}_{k}$ of its corresponding cluster.}}
\label{F6}
\end{figure}

Conditionally on $\hat{C}_i=k$, agency $i$ is associated with the cross-sell strategies suggested by $\hat{\bar{\boldsymbol{\pi}}}_k$. Hence, in ensuring accurate targeting, it is important to assess to what extent the estimated cluster-specific co-subscription probabilities adequately characterize the observed networks. Figure \ref{F6} displays the box plot  of the areas under the ROC curve (AUC), with each $\mbox{AUC}_i$, $i=1, \ldots, n$ obtained using the simulated  $\mathcal{L}(A_{i})$ and the estimated co-subscription probability vector $\hat{\bar{\boldsymbol  \pi}}_{k}$ of its corresponding cluster. Most of the co-subscription networks are adequately characterized by the co-subscription probability vectors specific to their clusters, with almost all the AUCs greater than $0.75$. Decreased predictive performance is obtained for a small set of agencies highlighted by dots. For such agencies, the company may devise ad-hoc cross-sell strategies based on the posterior mean of their agency-specific edge probability vector ${\boldsymbol \pi}_i$ rather than considering the cross-selling suggested to the clusters they belong to. 

To evaluate the fit with respect to the mono-product data, we consider the standardized $L_1$ distance between observed and estimated product frequencies $\epsilon_{i}=\sum_{v=1}^V| n_{iv}/n_i - \hat{p}_{kv} |/V$ with $k$ the cluster to which agency $i$ is allocated. In our simulation the maximum of these quantities is $\mbox{max}(\epsilon_{1}, \ldots, \epsilon_{n})=0.013$ meaning that mono-product choices in each cluster $k$ are adequately characterized by $\hat{{\bf p}}_k$.

\section{Cross-selling in the Italian insurance market}
\label{sec_5}
We apply the model outlined in Section \ref{sec_2} to our motivating business intelligence data set  described in Section \ref{sec_1}, which comprises mono-product choice data and co-subscription networks for $n=130$ agencies -- within the same company -- selling $V=15$ different insurance products. Posterior computation uses the same settings as in the simulation study. Also in this case we obtain convergence, good mixing and no issues of label switching. Similarly to the simulation study, clustering of agencies and the associated cross-sell strategies along with their performance indicators do not substantially change when considering $\alpha_c=0.5$, $\alpha_c=5$, $\alpha_c=10$, $\alpha_c=15$ and $\alpha_c=20$ instead of $\alpha_c=1$.

\begin{figure}[t]
\centering
\includegraphics[trim=0cm 0cm 0cm 0cm, clip=true,height=10cm, width=14.5cm]{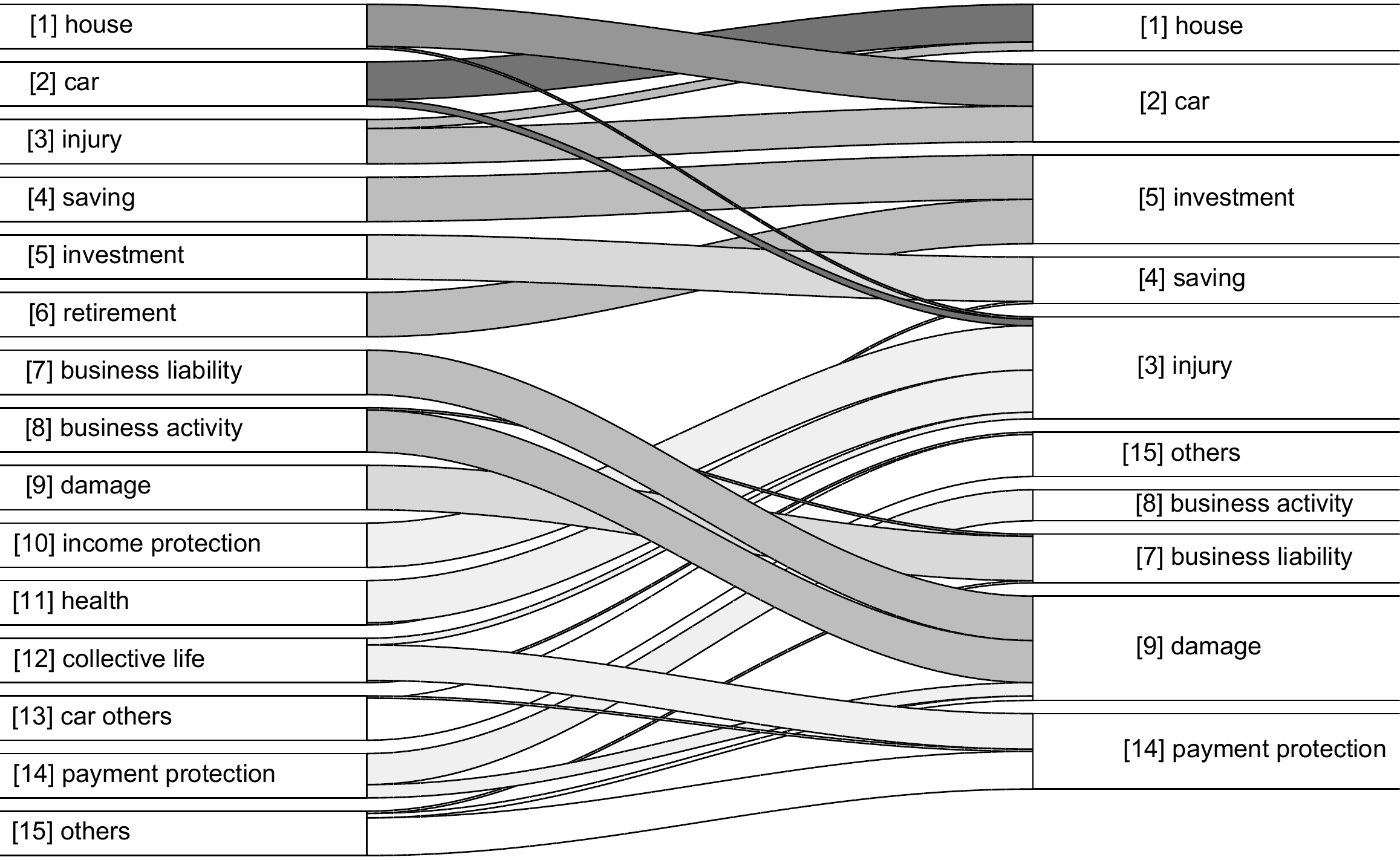}
\put (-417,290) {{{Mono-product segment}}}
\put (-107,290) {{{Best cross-sell offer}}}
\caption{\footnotesize{Flow chart summarizing the estimated cluster-specific cross-sell strategies $\hat{u}_{k1}, \ldots, \hat{u}_{kV}$ along with their performance indicators  $\hat{e}_{k1}, \ldots, \hat{e}_{kV}$ for the $k=1, \ldots, 20$ clusters in our application. The dimension of each edge connecting every product $v$ -- subscribed by current mono-product customers -- to a best offer is proportional to the total number of clusters for which that best offer coincides. The gray scale of the color for the edge starting from $v$, is proportional to its performance indicators averaged across clusters $\sum_{k=1}^{20} \hat{e}_{kv}/20$, $v=1, \ldots, 15$. The dimension of each right box is proportional to the  total number of different cross-sell strategies suggesting the corresponding product as the best offer.}}
\label{App_2}
\end{figure}

The posterior distribution for the cluster assignments suggests a total of $20$ clusters in our data. This is an appealing reduction of dimensionality in allowing the company to define $20$ sets of shared cross-sell strategies, instead of $n=130$ different campaigns. Figure \ref{App_2} provides a summarized overview of  our estimated cluster-specific cross-sell strategies $\hat{u}_{k1}, \ldots, \hat{u}_{kV}$ along with their performance indicators  $\hat{e}_{k1}, \ldots, \hat{e}_{kV}$, $k=1, \ldots, 20$. As we can clearly notice, the different clusters are typically characterized by similar cross-sell strategies, highlighting minor differences in mono- and multi-product customer profiles across different clusters of agencies. This is a reasonable insight provided that the focus of our study is on agencies operating in similar markets within the same insurance company. 

Accurate modeling of the cluster-specific probabilistic generative mechanisms for the co-subscription networks allows the definition of efficient sets of cluster-specific cross-sell strategies, offering to mono-product customers policies complementary to their current choice. According to Figure \ref{App_2}, family/individual-type policies, such as insurance for houses, cars, injuries, savings, investments and retirement plans, are typically associated with cross-sell offers from the same basket. The same holds for business-type policies covering business liability insurance, business activity insurance, compensatory damages and payment protection. Deepening the analysis of the Figure \ref{App_2}, popular best offers for cross-selling comprise insurance on investments,  insurance on injuries and compensatory damages. The former reasonably attracts mono-product customers currently subscribed to policies on savings and retirements plans, while the latter is a convenient offer for customers who are currently subscribed to business-type insurance products. Finally insurance on injuries is a profitable offer for mono-product customers associated with  health-type policies, such as medical health coverage and income protection insurance. The latter guarantees benefits to policy holders who are unable to work due to illness.

Our results provide consistent findings highlighting the accuracy of our procedure for profiling and cross-selling. This is further confirmed by the analysis of the performance indicators associated to each strategy. According to Figure \ref{App_2} house and car insurance are in general more effective in creating new multi-product customers. Beside being characterized by high co-subscription probabilities with other polices, these products are also highly populated by mono-product customers as house insurance represents a common policy for families and car insurance -- covering third-party liability policies -- is compulsory in Italy.

\begin{figure}[t]
\centering
\includegraphics[trim=0.2cm 0cm 0cm 0cm, clip=true,height=10.5cm, width=14cm]{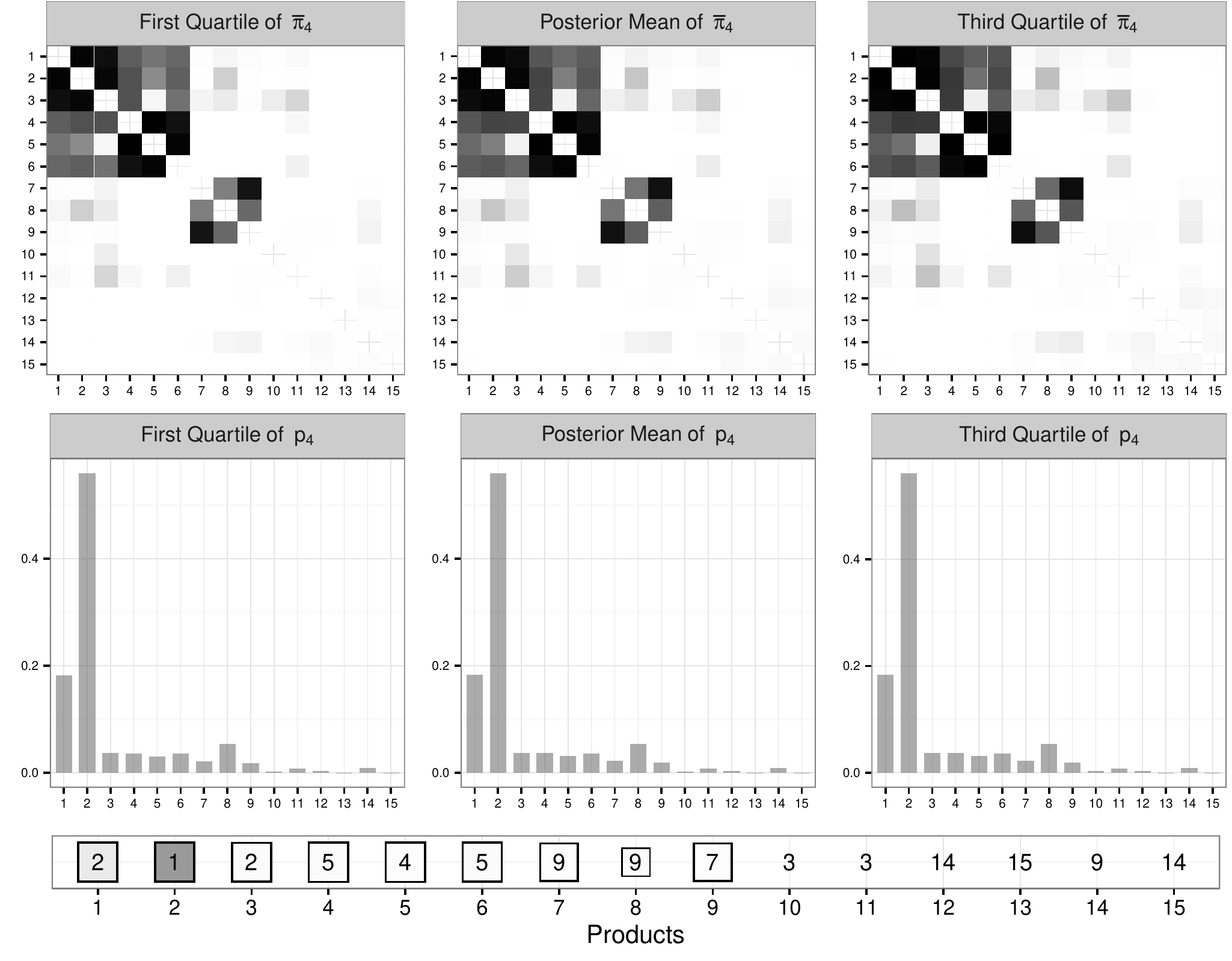}
\caption{\footnotesize{For the most populated cluster $k=4$. Upper panels: summary of the posterior distribution of the co-subscription probabilities ${\bar{\pi}}_{4l}$, $l=1, \ldots, V(V-1)/2$, rearranged in matrix form. Color goes from white to black as the probability goes from $0$ to $1$. Middle panels: summary of the posterior distribution of the mono-product choice probabilities $p_{4v}$, $v=1, \ldots, 15$. Lower panel: summary of the estimated cluster-specific cross-sell strategies $\hat{u}_{41}, \ldots, \hat{u}_{4V}$ along with their performance indicators  $\hat{e}_{41}, \ldots, \hat{e}_{4V}$.  The number in cell $v$ corresponds to the best offer $ \hat{u}_{4v}$. The dimension of each square $v$ is proportional to $\mbox{max}\{ \hat{\mbox{pr}}(\mathcal{A}_{4[vu]}=1): u \neq v\}$, while the gray scale of the color is proportional to the corresponding estimated performance $\hat{e}_{4v}$. To facilitate graphical analysis only squares associated to strategies with $\mbox{max}\{ \hat{\mbox{pr}}(\mathcal{A}_{4[vu]}=1): u \neq v\}>0.5$ are represented. }}
\label{App_1}
\end{figure}

Figure \ref{App_1} provides further clarifications of our findings by focusing on the posterior summaries for the key quantities ${\bar{\boldsymbol \pi}}_{4}$ and ${\bf p}_4$ associated to the most populated cluster $k=4$, comprising 40 agencies. According to Figure \ref{App_1},  our procedure provides a good joint representation of the different mono- and multi-product sources of variability, while confirming previous insights on mono and multi-product customer choices with respect to insurance products. Focusing on the posterior mean of ${\bar{\boldsymbol \pi}}_{4}$, we notice two communities of products characterized by high co-subscription probabilities among polices in the same community and comparatively lower co-subscription probabilities between policies in different communities. Consistently with results in Figure \ref{App_2}, the first group comprises the family/individual-type policies previously discussed, while the second contains the business related ones. The remaining less common policies in customer choices are instead characterized by low co-subscription probabilities, and therefore are expected to suggest poor cross-sell strategies. In addition the family/individual-type policies are further characterized by two sub-communities. The first comprises house insurance, car insurance and insurance on injuries, while the second contains insurance on savings, insurance on investments and retirement plans. This is a consistent finding, as the two sub-communities properly characterize non-life and life insurances policies, respectively.

\begin{figure}[t]
\centering
\includegraphics[height=7.8cm, width=15cm]{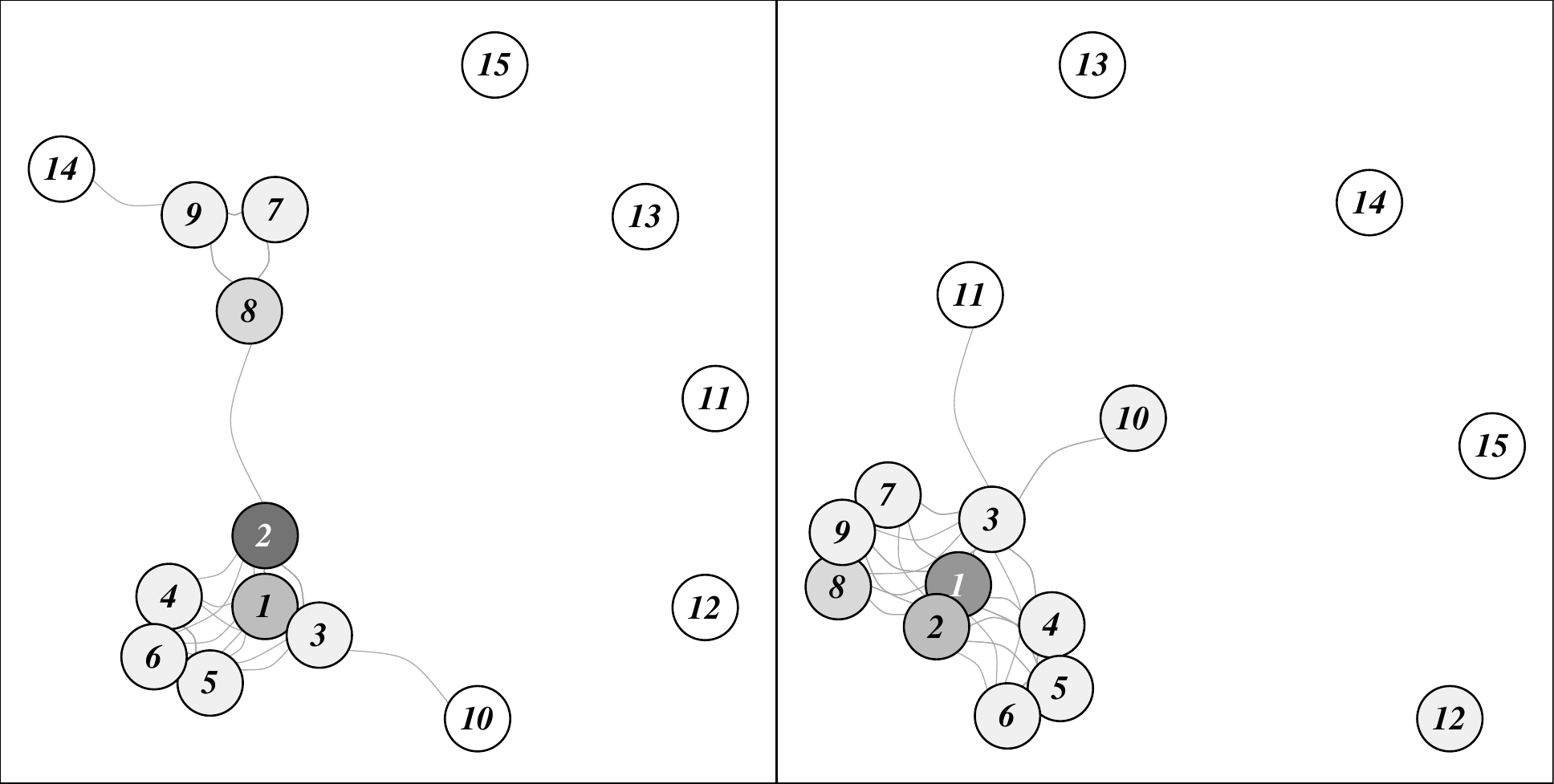}
\put (-420,210) {{{Cluster $3$}}}
\put (-205,210) {{{Cluster $14$}}}
\caption{\footnotesize{For two representative clusters, weighted network visualization with weights given by the posterior mean $\hat{\bar{\boldsymbol  \pi}}_{k}$ of their associated co-subscription probability vectors ${\bar{\boldsymbol  \pi}}_{k}$. Nodes positions are obtained applying the \cite{fru_1991} force-directed placement algorithm, while the gray scale of the color of each node $v$ is proportional to the corresponding estimated performance  indicator $\hat{e}_{kv}$, $v=1, \ldots, 15$.}}
\label{App_3}
\end{figure}

The posterior mean of ${\bf p}_{4}$ in Figure  \ref{App_1} confirms the previous comments on the performance indicators. Consistently with Figure  \ref{App_2}, house and car insurance policies are -- in fact -- the two products with the highest mono-product choice probabilities. Therefore, these policies are highly populated by mono-product customers. As a result, cross-sell strategies targeting these two segments have a high ceiling on effectiveness. Consistently  with these results, the cross-sell strategies for cluster $4$, along with their performance indicators, are mostly in line with those discussed in Figure  \ref{App_2}.

Posterior quartiles in Figure   \ref{App_1} additionally highlight how accounting for network structure and borrowing information  within and between clusters provide posterior distributions efficiently concentrated around our estimates. Posterior concentration is more evident for the mono-product choice probabilities. This result is not surprising provided that data on mono-product customers for agencies in cluster $4$ jointly contribute to the posterior of parameters in ${\bf p}_4$, with the average number of mono-product customers in each agency being $\approx 5{,}500$. 

Figure \ref{App_3} displays the estimated $\hat{\bar{{\boldsymbol\pi}}}_k$ and $\hat{{\bf p}}_k$, for two additional representative clusters. As we can notice, although mono- and multi-product customers of agencies in different clusters are characterized by a very similar behavior, our flexible procedure is able to capture subtle differences in the cluster-specific co-subscription probabilities and mono-product customer choices. The estimated co-subscription probabilities in cluster $3$ confirm the two community structure previously discussed for cluster $4$, while further highlighting the non-life and life sub-communities within the family/individual-type policies. Similarly to cluster $4$, mono-product customers in cluster $3$ have relatively high preferences for house insurance and car insurance. However we found the latter  slightly less populated in cluster $3$ in favor of house insurance policies, modifying the corresponding performance indicators. Correctly identifying differences in ${\bf p}_{k}$ is a key to evaluate and rank the cross-sell strategies $\hat{u}_{k1}, \ldots, \hat{u}_{kV}$ according to their performance indicators $\hat{e}_{k1}, \ldots, \hat{e}_{kV}$. 

The co-subscription network in cluster $14$ displays instead more evident differences, with less separation between the two communities characterizing family/individual-type policies and business ones. Hence, this cluster may refer to agencies dealing with customers who are motivated to co-subscribe policies both for their business activity and for their private coverage. Moreover -- differently from clusters $3$ and $4$ -- the mono-product preferences towards house insurance are higher than car insurance in cluster $14$. Therefore, targeting the mono-product customers currently subscribed to house insurance -- instead of the car insurance -- may be more effective at increasing the number of multi-product customers in agencies within cluster $14$.

\begin{figure}
\centering
\includegraphics[height=5.3cm, width=13cm]{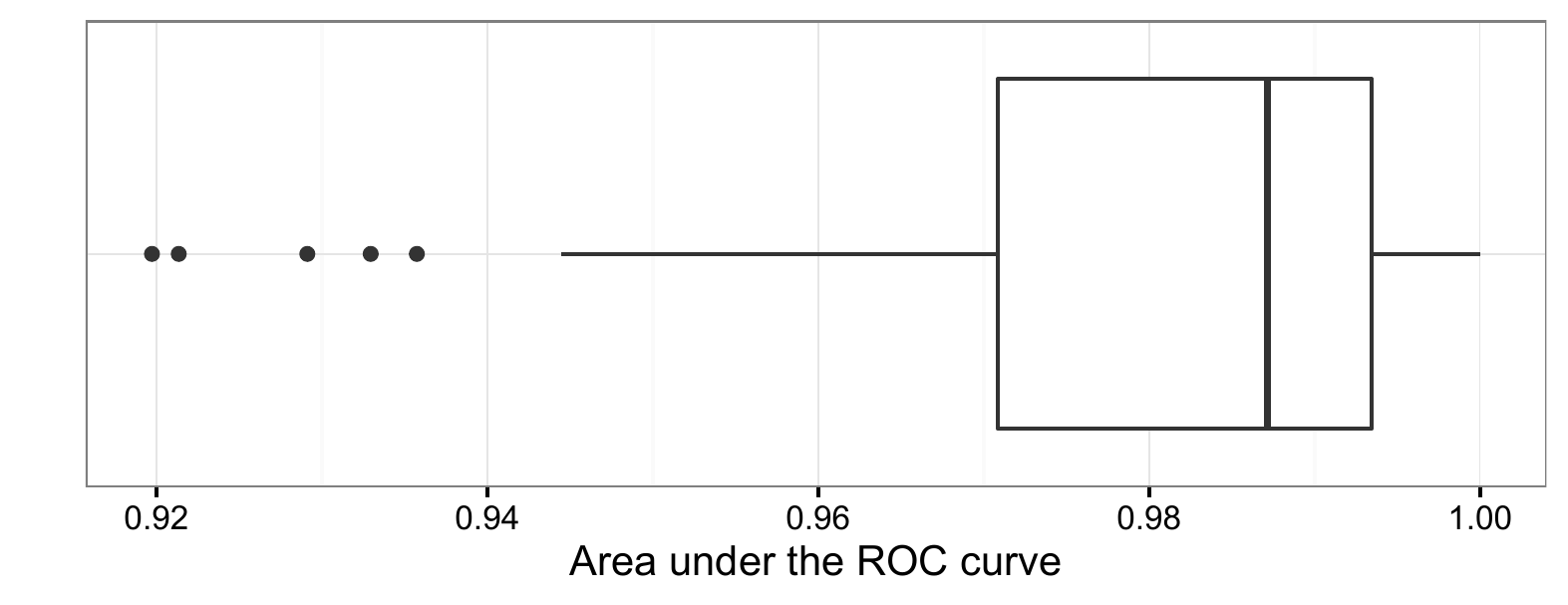}
\caption{\footnotesize{Box plot of the areas under the ROC curve $\mbox{AUC}_1, \ldots, \mbox{AUC}_{130}$. Each $\mbox{AUC}_i$ is  obtained using the observed $\mathcal{L}(A_{i})$ and the estimated co-subscription probability vector $\hat{\bar{\boldsymbol  \pi}}_{k}$ of its corresponding cluster.}}
\label{AUC_app}
\end{figure}

According to the box plot of the AUCs in Figure  \ref{AUC_app}, we do a very good job in characterizing the observed co-subscription networks $\mathcal{L}(A_i)$, $i=1, \ldots, 130$ considering the co-subscription probability vectors $\hat{\bar{\boldsymbol \pi}}_{k}$, $k=1, \ldots, 20$ specific to their clusters. All the AUCs are greater than $0.9$, with few agencies displaying a lower performance. For such agencies, the company may devise cross-sell strategies based on their specific $\hat{\boldsymbol \pi}_i$.  Our estimates provide also a good  fit with respect to the mono-product choice data with the maximum of the standardized $L_1$ distances between observed and estimated product frequencies for each agency being $\mbox{max}(\epsilon_{1}, \ldots, \epsilon_{130})=0.055$.

\section{Discussion}
\label{sec_6}
Motivated by a complex business intelligence problem for targeted advertising of cross-sell strategies in different agencies, we developed a flexible joint model for mixed domain data including mono-product customer choices and co-subscription networks measuring multiple purchasing behavior. Our procedure defines shared sets of cross-sell strategies by effectively clustering agencies characterized by comparable customer behaviors. Each segment is carefully profiled by  modeling mono-product customer choices and co-subscription networks via a cluster-dependent mixture of latent eigenmodels. Exploiting such estimates, we construct cluster-specific sets of cross-sell strategies informing for each product $v$ which additional product $u \neq v$ should be offered to obtain the highest probability of a co-subscription by a mono-product customer subscribed to $v$. We evaluate the effectiveness of each strategy via performance indicators accounting also for mono-product customer choice data. We provide straightforward algorithms for posterior computation. The application to multiple agencies from an Italian insurance company highlights the key benefits of our model in providing interesting insights and effective strategies. 

Although we focus on cross-sell strategies offering a single additional product, our procedure can be easily generalized to multi-offer cross-sell strategies, advertising to mono-product customers currently subscribed to $v$ in agencies within cluster $k$ the additional set of products ${\bf u}_{kv}=\{u_{kv1}, \ldots, u_{kvM}\}=\mbox{argmax}_{u_1, \ldots, u_M}\{ {\mbox{pr}}(\mathcal{A}_{k[vu_1]}=1,  \ldots, \mathcal{A}_{k[vu_M]}=1): u_1 \neq \ldots \neq u_M \neq v\}$. As our representation \eqref{eq_2}--\eqref{eq_3} induces a probability mass function on the entire co-subscription network, the quantity ${\mbox{pr}}(\mathcal{A}_{k[vu_1]}=1,  \ldots, \mathcal{A}_{k[vu_M]}=1)$ can be easily derived by marginalizing out in \eqref{eq_2} all the pairs of products not entering the previous joint probability. It is also worth considering further research including additional information, such as costs of the strategies and product prices. Such quantities impact on the definition and evaluation of the strategies along with mono- and multi-product choices. Our model still provides a valuable building block in estimating such quantities.

Although we focus on cross-selling in business intelligence, our model has a broad range of  applications. For example, our procedure can provide key information on efficient allocation of resources in public services based on mono- and multi-service citizen data from different cities or states. 
\bibliographystyle{rss}
\bibliography{daniele}

\end{document}